\documentclass{elsart}

\usepackage{epsfig}
\newcommand{\be}{\begin{equation}}
\newcommand{\ee}{\end{equation}}
\newcommand{\bea}{\begin{eqnarray}}
\newcommand{\eea}{\end{eqnarray}}

\newcommand{\Fritiof}{\textsc{Fritiof}}

\newcommand{\UNIT}[1]{\mbox{$\,{\rm #1}$}}

\newcommand{\GeV}{\UNIT{GeV}}

\begin{document}

\begin{frontmatter}
\title{Dilepton production and off-shell transport dynamics at SIS energies}
\author[FIAS]{E.L. Bratkovskaya,\corauthref{cor1}}
\ead{Elena.Bratkovskaya@th.physik.uni-frankfurt.de}
\corauth[cor1]{corresponding author}
\author[unig]{W.~Cassing,}
\address[FIAS]{Frankfurt Institute for Advanced Studies,
 Johann Wolfgang Goethe University,
  Ruth-Moufang-Str. 1,
 60438 Frankfurt am Main,
 Germany}
\address[unig]{Institut f\"ur Theoretische Physik, %
  Universit\"at Giessen,
  Heinrich-Buff-Ring 16,
  D-35392 Giessen, %
  Germany}

\begin{abstract}
Dilepton production in nucleus-nucleus collisions at 1-2
A$\cdot$GeV as well as in elementary $pp$ and $pd$ reactions is
studied within the microscopic HSD transport approach which
includes the off-shell dynamics of vector mesons explicitly. The
study addresses additionally the production of $\pi^0$ and $\eta$
mesons since their Dalitz decays provide a sizeable contribution
to the dilepton invariant mass spectra up to about 0.5 GeV. Our
transport results agree with the TAPS experimental data on $\pi$
and $\eta$ multiplicities in $^{12}C + ^{12}C$ collisions from 0.8
to 2 A GeV. We find that the 'DLS-puzzle' - which stands for a
theoretical underestimation of the $e^+e^-$ yield in the mass
range from 0.2 to 0.5 GeV in $^{12}C + ^{12}C$ and $^{40}Ca +
^{40}Ca$ collisions - may be solved when incorporating a stronger
bremsstrahlung contribution in line with recent OBE calculations.
Moreover, the HSD results with  'enhanced' bremsstrahlung cross
sections agree very well  with the HADES experimental data for the
dilepton mass spectra for $^{12}C + ^{12}C$ at 1 and 2
A$\cdot$GeV, especially when including a collisional broadening in
the vector-meson spectral functions. Detailed predictions for
dilepton spectra from $pp$ and $pn/pd$ reactions at 1.25 GeV, 2.2
GeV and 3.5 GeV are presented which will allow to verify/falsify
the larger bremsstrahlung contributions from the experimental side
in the near future.

\end{abstract}


\end{frontmatter}

\newpage

\section{Introduction}

The theory of quantum-chromo-dynamics (QCD) describes hadrons as
many-body bound or resonant states of partonic constituents. While
the  properties of hadrons are rather well known in free space
(embedded in a nonperturbative QCD vacuum) the mass and lifetimes
of hadrons in a baryonic and/or mesonic environment are subject of
current research in order to achieve a better understanding of the
strong interaction and the nature of confinement. In this context
the modification of hadron properties in nuclear matter is of
fundamental interest (cf. Refs.
\cite{BrownRho,H&L92,asakawa,Chanfray,Asakawa93,Herrmann,Shakin94,Rapp,Klingl96,Friman,RappNPA,Peters}
for the leading concepts) since QCD sum
rules \cite{H&L92,Asakawa93,Leupold} as well as QCD inspired
effective Lagrangian models
\cite{BrownRho,Chanfray,Herrmann,Shakin94,Rapp,Klingl96,Friman,RappNPA,Peters}
predict  significant changes e.g. of the vector mesons ($\rho$,
$\omega$ and $\phi$) with the nuclear density $\rho_N$ and/or
temperature $T$.

A more direct evidence for the modification of vector mesons has been
obtained from the enhanced production of lepton pairs above known
sources in nucleus-nucleus collisions at SPS energies
\cite{CERES,HELIOS}. As proposed by Li, Ko, and Brown \cite{Li} and Ko
et al. \cite{Li96} the observed enhancement in the invariant mass range
$0.3 \leq M \leq 0.7$ GeV might be due to a shift of the $\rho$-meson
mass following Brown/Rho scaling \cite{BrownRho} or the Hatsuda and Lee
sum rule prediction~\cite{H&L92}.  The microscopic transport studies in
Refs. \cite{Cass95C,Cass96H,Brat97,CBRep98,Ernst} for these systems
have given support for this interpretation \cite{Li,Li96,Ko93,Ko95}. On
the other hand also more conventional approaches that describe a
melting of the $\rho$-meson in the medium due to the strong hadronic
coupling (along the lines of
Refs.~\cite{asakawa,Chanfray,Herrmann,Rapp,Peters}) have been found to
be compatible with the early CERES data
\cite{Rapp,Cass95C,CBRW97,rapp4,rapp5}.

This ambiguous situation has been clarified to some extent by the NA60
Collaboration since the invariant mass spectra for $\mu^+\mu^-$ pairs
from In+In collisions at 158 A$\cdot$GeV clearly favored the 'melting
$\rho$' scenario \cite{NA60}. Also the latest data from the CERES
Collaboration (with enhanced mass resolution) \cite{ceres2} show a
preference for the 'melting $\rho$' scenario which is in line with more
recent theoretical studies on dilepton production at SPS energies
\cite{rapp1,rapp2,rapp3,gale1,gale2,renk12,renk3,Dusling}.

Dileptons have also been measured in heavy-ion
collisions at the BEVALAC by the DLS Collaboration \cite{DLSold,DLSnew}
at incident energies that are two orders-of-magnitude lower than that
at the SPS. The first published spectra at 1 A$\cdot$GeV \cite{DLSold}
(based on a limi\-ted data set) have been consistent with the results
from transport model calculations \cite{Xiong90,Wolf90,Gudima,BCMas96}
that include $pn$ bremsstrahlung, $\pi^0$, $\eta$ and $\Delta$ Dalitz
decay and pion-pion annihilation.  However, in 1997 the DLS
Collaboration released a new set of data \cite{DLSnew} based on the
full data sample and an improved analysis, which showed a considerable
increase in the dilepton yield:  more than a factor of five above the
early DLS data \cite{DLSold} and the corresponding theoretical results
\cite{Xiong90,Wolf90,Gudima,BCMas96}. This discrepancy remained in the
transport calculations even after including contributions from the
decay of $\rho$ and $\omega$ mesons that are produced directly from
nucleon-nucleon and pion-nucleon scattering in the early reaction phase
\cite{Ernst,BratRapp98}. With an in-medium $\rho$ spectral function, as
that used in Ref. \cite{CBRW97} for dilepton production from heavy-ion
collisions at SPS energies, dileptons from the decay of both directly
produced $\rho$'s and pion-pion annihilation have been considered, and
a factor of about two enhancement has been obtained in the theoretical
studies compared to the case of a free $\rho$-spectral function.
Furthermore, in Ref.  \cite{BratKo99} the alternative scenario of a
dropping $\rho$-meson mass and its influence on the properties of the
$N(1520)$ resonance has been investigated. Indeed, an incorporation of
such medium effects lead to an enhancement of the $\rho$-meson yield,
however, was not sufficient to explain the DLS data. Since  independent
transport calculations by HSD (BUU) and UrQMD  underestimated the DLS
data for $C+C$ and $Ca+Ca$ at 1 A GeV by roughly the same amount these
findings have led to the denotation 'DLS-puzzle' in 1999
\cite{Ernst,BratRapp98,BratKo99} which persists now by about a decade.

As has been shown in 2003 by the transport analysis of the T\"ubingen
group \cite{Fuchs03} also alternative scenarios for the in-medium
modification, i.e. a possible decoherence between the intermediate
mesonic states in the vector resonance decay, increases the dilepton
yield. However, still  the region about $M\simeq 0.3$ GeV was
underestimated whereas the yield in the vicinity of the $\rho$-meson
peak was overestimated (especially for $C+C$ collisons).  Thus there is
no consistent explanation for the DLS data so far and questions came up
about the validity of the DLS data or a possible misinterpretation of
the DLS acceptance cuts.

In order to address the in-medium modifications of vector-mesons from
the experimental side the HADES spectrometer has been built
\cite{HADES06} which allows to study $e^+e^-$ pair production in a much
wider acceptance region for elementary $pp$, $pd$ reactions as well as
$\pi A$, $p A$  or even $A A$ collisions up to about 8 A$\cdot$GeV.
During the construction phase of HADES there has been a lot of
theoretical activity providing various predictions for these reactions
at SIS energies
\cite{CBRep98,Brat_pA01,Brat_KEK02,Effe_piA,fuchs1,fuchs2,fuchs3,wolfg1,wolfg2}.
Meanwhile the HADES Collaboration has presented first spectra
\cite{hade1,HADES06,HADES07pt,HADES07} and the question is: what do these
data tell us?

The answer to the questions raised above is nontrivial due to the
nonequilibrium nature of these reactions and transport models have
to be incorporated to disentangle the various sources that
contribute to the final dilepton spectra seen experimentally. In
this study we aim at contributing to this task employing an
up-to-date relativistic transport model (HSD) that incorporates
the relevant off-shell dynamics of the vector mesons. Compared to
our earlier studies in Refs. \cite{CBRep98,CBRW97,BratRapp98,BratKo99}
a couple of extensions have been implemented  such as
\begin{itemize}
\item{off-shell dynamics for vector mesons - according to
Refs. \cite{Cass_off1,Cass_off2} - and an explicit
(nonperturbative) propagation of $\rho, \omega$ and $\phi$} also
at SIS energies
\item{extension of the LUND string model to include 'modified'
spectral functions for the resonances in the string decays}
\item{novel cross sections for $\eta$ production in line with more
recent experimental data}
\item{novel $NN$ bremsstrahlung cross sections according to
recent OBE calculations}
\item{an extended set of spectral functions for the vector mesons
$\rho, \omega, \phi$.}
\end{itemize}
We note that the first off-shell dynamical study for vector mesons
and dilepton decays has been presented in Ref. \cite{Brat_pA01}
for $pA$ reactions and been picked up later by the Rossendorf group
\cite{wolfg2} also incorporating the off-shell
equations from Refs. \cite{Cass_off1,Cass_off2}.
The off-shell dynamics is particularly important for resonances with a rather long
lifetime in vacuum but strongly decreasing lifetime in the nuclear
medium (especially $\omega$ and $\phi$ mesons).

The outline of the paper is as follows: After a brief description
of the transport model and its elementary channels in Section 2 we
will discuss $\pi^0$ and $\eta$ production in elementary as well
as $AA$ reactions in comparison to available data  (Section 3) in
order to have independent constraints on the leptonic sources.
Section 4 is devoted to dilepton production in $pp$ and $pd$
reactions in confrontation with spectra from the DLS while in
Section 5 a comprehensive study of $^{12}C+^{12}C$ and $^{40}Ca +
^{40}Ca$ collisions is performed employing different spectral
functions for the vector mesons. In addition we will present
predictions for future projects of the HADES Collaboration like
$Ar+KCl$ at 1.75 A$\cdot$GeV or $pp$, $pd$ reactions at 1.25 GeV,
2.2 GeV and 3.5 GeV. A summary and discussion closes this work in
Section 6.

\section{Description of the model}

Our analysis is carried out within the HSD transport model
\cite{Brat97,CBRep98,Ehehalt} - based on covariant self energies
for the baryons \cite{KWeber} - that has been used for the
description of $pA$ and $AA$ collisions from SIS to RHIC energies.
We recall that in the HSD approach nucleons, $\Delta$'s,
N$^*$(1440), N$^*$(1535), $\Lambda$, $\Sigma$ and $\Sigma^*$
hyperons, $\Xi$'s, $\Xi^*$'s and $\Omega$'s as well as their
antiparticles are included on the baryonic side whereas the $0^-$
and $1^-$ octet states are incorporated in the mesonic sector.
Inelastic baryon--baryon (and meson-baryon) collisions with energies
above $\sqrt s_{th}\simeq 2.6\GeV$ (and  $\sqrt s_{th}\simeq 2.3\GeV$) are
described by the \Fritiof{} string model \cite{FRITIOF} whereas low
energy hadron--hadron collisions are modeled in line with experimental
cross sections.

The dilepton production by a (baryonic or mesonic) resonance $R$
decay can be schematically presented in the following way:
\begin{eqnarray}
 BB &\to&R X   \label{chBBR} \\
 mB &\to&R X \label{chmBR} \\
      && R \to  e^+e^- X, \label{chRd} \\
      && R \to  m X, \ m\to e^+e^- X, \label{chRMd} \\
      && R \to  R^\prime X, \ R^\prime \to e^+e^- X, \label{chRprd}
\end{eqnarray}
i.e. in a first step a resonance $R$ might be produced in
baryon-baryon ($BB$) or meson-baryon ($mB$) collisions
(\ref{chBBR}), (\ref{chmBR}). Then this resonance can couple to
dileptons directly (\ref{chRd}) (e.g., Dalitz decay of the
$\Delta$ resonance: $\Delta \to e^+e^-N$) or decays to a meson $m$
(+ baryon) or in (\ref{chRMd})  produce dileptons via direct
decays ($\rho, \omega$) or Dalitz decays ($\pi^0, \eta, \omega$).
The resonance $R$ might also decay into another resonance
$R^\prime$  (\ref{chRprd}) which later produces dileptons via
Dalitz decay.  Note, that in the combined model the final
particles -- which couple to dileptons -- can be produced also via
non-resonant mechanisms, i.e. 'background' channels at low and
intermediate energies or string decay at high energies
\cite{Falter}.

The electromagnetic part of all conventional dilepton sources  --
$\pi^0, \eta, \omega$  Dalitz decays, direct decay of vector
mesons $\rho, \omega$ and $\phi$ -- are treated as described in
detail in Ref.~\cite{BCM00SIS}-- where dilepton production in $pp$
and $pd$ reactions has been studied. Modifications -- relative to
Ref.~\cite{BCM00SIS} -- are related to the Dalitz decay of
baryonic resonances and especially the strength of the $pp$ and
$pn$ bremsstrahlung since calculations by Kaptari and
K\"ampfer in 2006 \cite{Kaptari} indicated that the latter channels might
have been severely underestimated in our previous studies on
dilepton production at SIS energies. In detail: \\ -- For the
Dalitz decays of the baryonic resonances we adopt the
parametrizations from Ernst et al. \cite{Ernst} (eqs. (9) to (13))
which are also incorporated in the PLUTO simulation program of the
HADES Collaboration \cite{PLUTO}.\\ -- For the bremsstrahlung
channels in $pp$ and $pn$ reactions we adjust our previous
expressions in order to match the recent results from the OBE model
calculations in Ref. \cite{Kaptari} (cf. Section 2.5).

\subsection{Vector-meson spectral functions}

\begin{figure}[t]
\centerline{{\psfig{figure=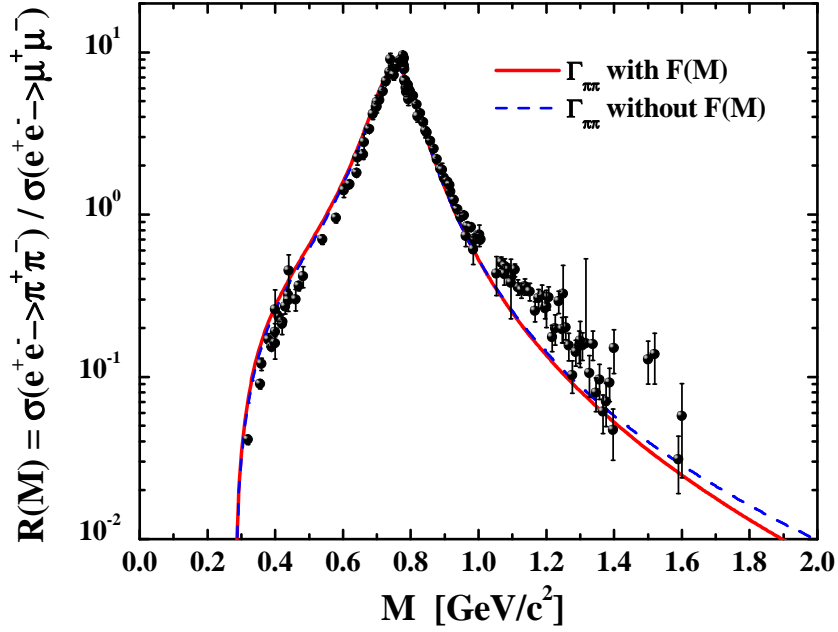,width=11cm}}}
\caption{The ratio $R(M)=\sigma(e^+e^-\to \pi^+\pi^-)/
\sigma(e^+e^-\to \mu^+\mu^-)$. The solid (red) line corresponds to the
result with the formfactor $F(M)$ (\ref{Frapp}) while the dashed (blue) line
is without $F(M)$.  The compilation of the experimental data is
taken from Ref. \protect\cite{expFPI}.}
\label{Fpi}
\end{figure}

In order to explore the influence of in-medium effects on the
vector-meson spectral functions we incorporate the effect of
collisional broadening (as in Refs. \cite{BratKo99,GKC97}), i.e.
the vector meson width has been implemented as:
\begin{eqnarray}
\Gamma^*_V(M,|\vec p|,\rho_N)=\Gamma_V(M) + \Gamma_{coll}(M,|\vec
p|,\rho_N) . \label{gammas}
\end{eqnarray}
Here $\Gamma_V(M)$ is the total width of the vector mesons ($V=\rho,\omega$)
in the vacuum. For the $\rho$ meson we use
\begin{eqnarray}
\Gamma_\rho(M) &\simeq& \Gamma_{\rho\to\pi\pi}(M) =  \Gamma_0
\left(M_0\over M\right)^2 \left(q\over q_0\right)^3 \ F(M)
\label{Widthrho} \\
&& q = {(M^2-4m_\pi^2)^{1/2}\over 2}, \
  \ q_0 = {(M_0^2-4m_\pi^2)^{1/2}\over 2}. \nonumber
\end{eqnarray}
In (\ref{Widthrho}) $M_0$ is the vacuum pole mass of the vector meson
spectral function, $F(M)$ is a formfactor taken from Ref. \cite{Rapp} as
\begin{eqnarray}
F(M)={\left(2\Lambda^2 +M_0^2 \over  2\Lambda^2 + M^2 \right)^2}
\label{Frapp}\end{eqnarray}
with a cut-off parameter $\Lambda=3.1$~GeV. This formfactor was
introduced in Ref.  \cite{Rapp} in order to describe the $e^+e^-$
experimental data with better accuracy.
In Fig. \ref{Fpi} we show the ratio $R(M)=\sigma(e^+e^-\to
\pi^+\pi^-)/\sigma(e^+e^-\to \mu^+\mu^-)$ which is proportional to the
pion formfactor $F_\pi(M)$ except for a kinematical factor.  The solid
line corresponds to the result with the formfactor $F(M)$ (\ref{Frapp})
while the dashed line is without $F(M)$.  The compilation of the
experimental data is taken from Ref.  \cite{expFPI}. As seen from Fig.
\ref{Fpi} the effect of the formfactor $F(M)$ (\ref{Frapp}) is visible
only at high $M$.

\noindent
For the $\omega$ meson a constant total vacuum width is used:
$\Gamma_\omega\equiv \Gamma_\omega(M_0)$, since the $\omega$ is a
narrow resonance in vacuum.

The collisional width in (\ref{gammas}) is approximated as
\begin{eqnarray}
\Gamma_{coll}(M,|\vec p|,\rho_N) = \gamma \ \rho_N < v \
\sigma_{VN}^{tot} > \approx  \ \alpha_{coll} \
\frac{\rho_N}{\rho_0} . \label{dgamma}
\end{eqnarray}
Here $v=|{\vec p}|/E; \ {\vec p}, \ E$ are the
velocity, 3-momentum and energy of the vector meson in the rest frame
of the nucleon current and $\gamma^2=1/(1-v^2)$.
Furthermore, $\rho_N$ is the nuclear density and
$\sigma_{VN}^{tot}$ the meson-nucleon total cross section.

In order to simplify the actual calculations for dilepton production
the coefficient $\alpha_{coll}$ has been extracted in the HSD transport
calculations from the vector-meson collision rate in $C+C$ and $Ca+Ca$
reactions (at 1 and 2 A$\cdot$ GeV) as a function of the density
$\rho_N$. In case of the $\rho$ meson the collision rate is dominated
by the absorption channels
$\rho N \rightarrow \pi N$ or $\rho N \rightarrow \Delta \pi
\rightarrow \pi \pi N$
at SIS energies. Also the reactions $\rho +\pi \leftrightarrow a_1$ are
incorporated but of minor importance here.
The numerical results for $\Gamma_{coll}(\rho_N)$ then have
been divided by $\rho_N/\rho_0$ to fix the coefficient $\alpha_{coll}$
in (\ref{dgamma}).  We obtain  $\alpha_{coll} \approx 150$~MeV for the
$\rho$ and $\alpha_{coll} \approx 70$~MeV for $\omega$ mesons which are
consistent with Ref. \cite{Metag07}.  In this way the average effects
of collisional broadening are incorporated in accordance with the
transport calculations and allow for an explicit
representation of the vector-meson spectral functions versus the
nuclear density (see below). We point out, however, that only the
average width of the vector mesons is consistent with the
dynamical evolution of these particles in the transport calculations
and not the specific shape of  their spectral functions.

In order to explore the observable consequences of vector meson
mass shifts at finite nuclear density -- as indicated by the
CBELSA-TAPS data \cite{tapselsa} for the $\omega$ meson -- the
in-medium vector meson pole masses are modeled (optionally)
according to the Hatsuda and Lee \cite{H&L92} or Brown/Rho scaling
\cite{BrownRho} as
\begin{eqnarray}
\label{Brown}
M_0^*(\rho_N)= \frac{M_0} {\left(1 + \alpha {\rho_N / \rho_0}\right)},
\end{eqnarray}
where $\rho_N$ is the nuclear density at the resonance decay
position $\vec r$; $\rho_0 = 0.16 \ {\rm fm}^{-3}$ is the normal
nuclear density and $\alpha \simeq 0.16$ for the $\rho$ and
$\alpha \simeq 0.12$ for the $\omega$ meson \cite{Metag07}. The parametrization
(\ref{Brown}) may be employed also at much higher collision
energies (e.g. FAIR and SPS) and one does not have to introduce a
cut-off density in order to avoid negative pole masses. Note that
(\ref{Brown}) is uniquely fixed by the 'customary' expression
$M_0^*(\rho_N) \approx M_0 (1 - \alpha \rho_N/\rho_0)$ in the low
density regime.

\begin{figure}[t]
\phantom{a}\vspace*{5mm}
{\psfig{figure=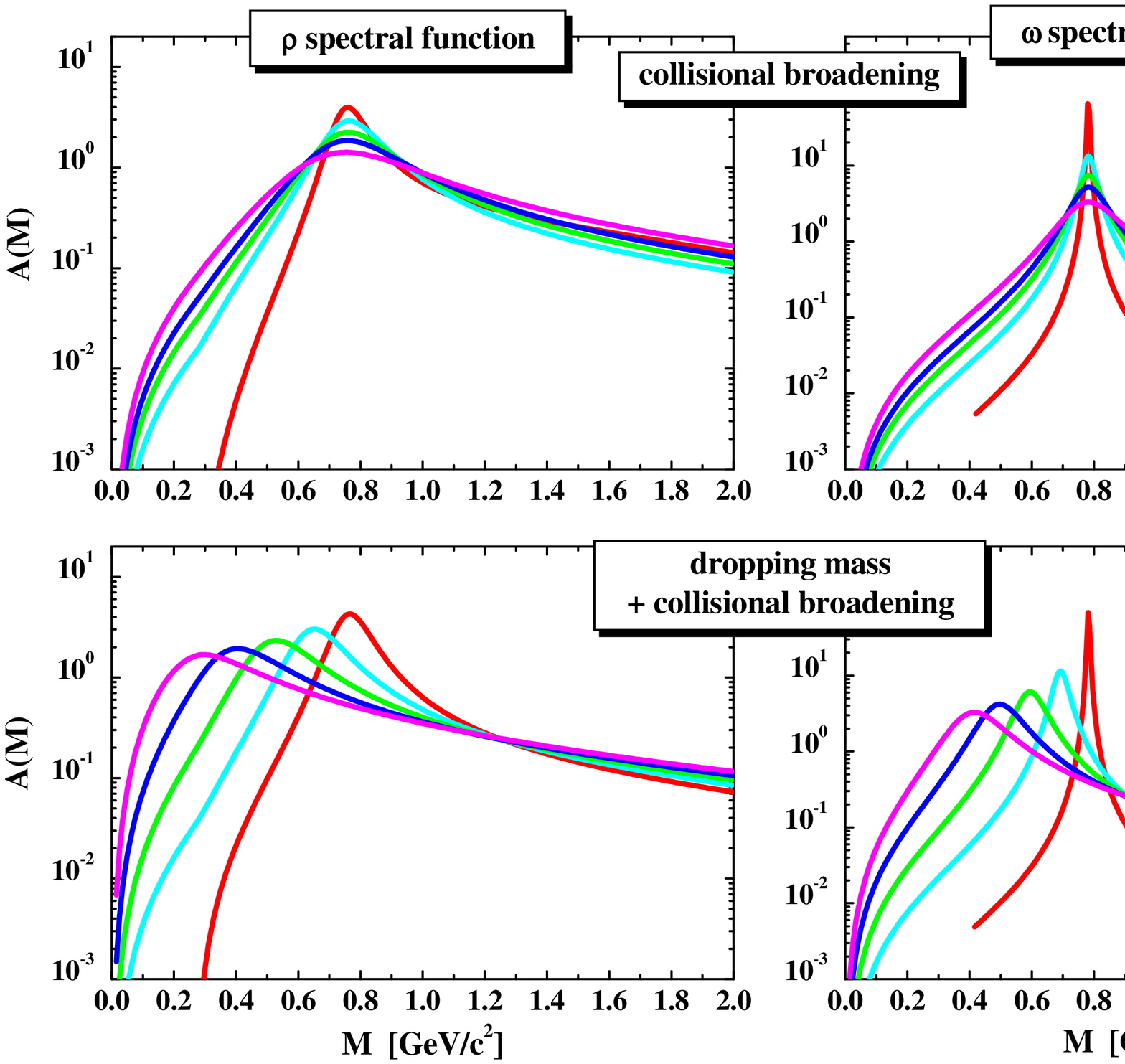,width=11cm}}
\caption{The spectral functions for the $\rho$ and $\omega$ meson
 in the case of the 'collisional broadening' scenario (upper part)
and the 'dropping mass + collisional broadening' scenario (lower
part) for nuclear densities of 0,1,2,3,5$\times\rho_0$ as
employed in the transport calculations (see text for details). }
\label{Fig0}
\end{figure}

The spectral function of the vector meson $V$ for the mass $M$
at baryon density $\rho_N$ is taken in the Breit-Wigner form:
\begin{eqnarray}
A_V(M,\rho_N) = C_1\cdot {2\over \pi} \ {M^2 \Gamma_V^*(M,\rho_N)
\over (M^2-M_{0}^{*^2}(\rho_N))^2 + (M {\Gamma_V^*(M,\rho_N)})^2}\
.
\label{spfunV}
\end{eqnarray}
The factor $C_1$ is fixed by the normalization condition for
arbitrary $\rho_N$:
\begin{eqnarray}
\int_{M_{min}}^{M_{lim}} A_V(M,\rho_N) dM =1,
\label{SFnorma}\end{eqnarray} where $M_{lim}=2$~GeV is chosen as
an upper limit for the numerical integration. The lower limit for
the vacuum spectral function corresponds to the two-pion decay,
$M_{min}=2 m_\pi$, whereas for the in-medium collisional broadening
case $M_{min}=2 m_e \to 0$ with $m_e$ denoting the electron mass.
$M_0^*$ is the pole mass of the vector meson spectral function which is
$M_0^*(\rho_N=0)=M_0$ in vacuum, however, may be  shifted in the medium
for the dropping mass scenario according to Eq.  (\ref{Brown}).

The resulting spectral functions for the $\rho$ and $\omega$ meson
are displayed in Fig. \ref{Fig0} for the case of 'collisional broadening'
(upper part) as well as for the 'dropping mass + collisional
broadening' scenario (lower part) for densities of 0,1,2,3,5 $\times
\rho_0$.  Note that in vacuum the hadronic widths vanish for the $\rho$
below the two-pion mass and for the $\omega$ below the three-pion mass.
With increasing nuclear density $\rho_N$ elastic and inleastic
interactions of the vector mesons shift strength to low invariant
masses. In the 'collisional broadening' scenario we find a dominant
enhancement of strength below the pole mass for the $\rho$ meson while
the $\omega$ meson spectral function is drastically enhanced in the
low- and high-mass region with density (on expense of the pole-mass
regime). In the 'dropping mass + collisional broadening' scenario both
vector mesons dominantly show a shift of strength to low invariant
masses with increasing $\rho_N$. Qualitatively similar pictures are
obtained for the $\phi$ meson but quantitatively smaller effects are
seen due to the lower effect of mass shifts and a substantially reduced
$\phi N$ cross section which is a consequence of the $s\bar{s}$
substructure of the $\phi$ meson. Since the $\phi$ dynamics turn out to
be of minor importance for the dilepton spectra to be discussed below
we discard an explicit representation.

The 'family' of spectral functions shown in Fig. 1 allows for a
sufficient flexibility with respect to the possible scenarios
outlined above. A comparison to dilepton data is expected to provide
further constraints on the possible realizations.

\subsection{Off-shell propagation}

The underlying concept of all 'on-shell' semi-classical transport
models (VUU, BUU, QMD, UrQMD etc.) is based on a Monte-Carlo solution of
the transport equations for the phase-space densities of hadrons
which are obtained in first order gradient expansion of the Wigner
transformed Kadanoff-Baym equations using the quasiparticle
approximation (QPA). The latter (QPA) assumes that the particles
have  narrow (i.e. $\delta$-like) spectral functions which don't
change their shape during the propagation. This concept works
sufficiently well for weakly interacting particles with a very
long lifetime but becomes inadequate for short-lived (broad)
resonance states of high collision rate.

A first attempt to dynamically account for in-medium effects of
vector mesons (i.e. a broadening of their spectral function) using
the standard on-shell transport model showed a couple of severe
problems (cf. the study by Effenberger et al. in Ref.
\cite{Effe}). The most serious one is that e.g. a $\rho$ meson
with a broad spectral function in the medium does not regain its
vacuum spectral shape when propagating out of the medium
\cite{Effe}. The reason is that dynamical changes of the spectral
function during propagation are  only included by explicit
collisions with other particles in conventional on-shell models
(cf. the discussion in context with Fig. 2 below).

In order to allow for a proper phase-space description of broad
resonances with varying spectral functions the off-shell transport
dynamics (OSTD) has been developed in Refs.
\cite{Cass_off1,Cass_off2}. The main idea of the OSTD is to
discard the quasiparticle approximation in the first order
gradient expansion of the Wigner transformed Kadanoff-Baym
equations. This leads to generalized transport equations which
contain an additional 'backflow' term that vanishes in the
on-shell QPA, however,  survives in the off-shell case and becomes
very important for the dynamics of broad resonances
\cite{Cass_off1,Cass_off2} (see below). A first application of the OSTD to
dilepton production has been presented already in 2001 by one of
the authors \cite{Brat_pA01}. It has been shown that using OSTD
the broad spectral function of vector mesons merges to the vacuum
shape by propagation out of the dense medium.

Here we briefly describe the implementation of OSTD in the HSD
model: A vector meson $V$ - created at some space-time point $x$
at density $\rho_N$ - is distributed in mass according to its
spectral function $A_V(M,\rho_N)$ (and the available phase space).
While propagating through the nuclear medium the total width of
the vector meson $\Gamma_V^* (M,\rho_N)$ (\ref{gammas}) changes
dynamical and its spectral function (\ref{spfunV}) is also
modified according to the real part of the vector meson self
energy $Re \Sigma^{ret}$ as well as by the imaginary part of the
self energy ($\Gamma_V^*\simeq -Im \Sigma^{ret}/M$) following
\begin{eqnarray}
A_V(M,\rho_N) = C_1\cdot {2\over \pi} \ {M^2 \Gamma_V^*
\over (M^2-M_0^2- Re \Sigma^{ret})^2 + (M {\Gamma_V^*})^2},
\label{spfun}
\end{eqnarray}
which is the in-medium form for a scalar boson spectral function.
The spectral function has to change during propagation
and to merge the vacuum spectral function when propagating out of
the medium.

In the OSTD the general off-shell equations of motion  for test
particles  with momentum $\vec P_i$, energy $\varepsilon_i$ at position
$\vec X_i$ read
\cite{Cass_off1,Cass_off2}
\begin{eqnarray}
\label{eomr} \frac{d {\vec X}_i}{dt} \! & = & \frac{1}{1-C_{(i)}}
\  \frac{1}{2 \varepsilon_i} \: \left[ \, 2  {\vec P}_i  +  {\vec
\nabla}_{P_i} \, Re \Sigma^{ret}_{(i)}  +  \frac{ \varepsilon_i^2
- {\vec P}_i^2 - M_0^2 - Re \Sigma^{ret}_{(i)}}{{\tilde
\Gamma}_{(i)}} \: {\vec \nabla}_{P_i} \, {\tilde \Gamma}_{(i)} \:
\right],
\\[0.3cm]
\label{eomp} \frac{d {\vec P}_i}{d t} \! & = & -
\frac{1}{1-C_{(i)}}\ \frac{1}{2 \varepsilon_{i}} \: \left[ {\vec
\nabla}_{X_i} \, Re \Sigma^{ret}_i \: + \: \frac{\varepsilon_i^2 -
{\vec P}_i^2 - M_0^{2} - Re \Sigma^{ret}_{(i)}}{{\tilde
\Gamma}_{(i)}} \: {\vec \nabla}_{X_i} \, {\tilde \Gamma}_{(i)} \:
\right],
\\[0.3cm]
\label{eome} \frac{d \varepsilon_i}{d t} & = &
\frac{1}{1-C_{(i)}}\ \frac{1}{2 \varepsilon_i} \: \left[
\frac{\partial Re \Sigma^{ret}_{(i)}}{\partial t} \: + \:
\frac{\varepsilon_i^2 - {\vec P}_i^2 - M_0^{2} - Re
\Sigma^{ret}_{(i)}}{{\tilde \Gamma}_{(i)}} \: \frac{\partial
{\tilde \Gamma}_{(i)}}{\partial t} \right],
\end{eqnarray}
where the notation $F_{(i)}$ implies that the function is taken at
the coordinates of the test particle, i.e. $F_{(i)} \equiv
F(t,\vec{X}_{i}(t),\vec{P}_{i}(t),\varepsilon_{i}(t))$. In Eqs.
(\ref{eomr})-(\ref{eome}) $Re \Sigma^{ret}$ denotes the real part
of the retarded self energy while ${\tilde \Gamma} = -Im
\Sigma^{ret} $ stands for the (negative) imaginary part in short-hand
notation. Apart from the propagation in the real potential $\sim Re
\Sigma^{ret}/2\varepsilon$ the equations (\ref{eomr}) --
(\ref{eome}) include the dynamical changes due to the imaginary
part of the self energy $Im \Sigma^{ret} \sim - M \Gamma_{V}^*$
with $\Gamma_V^*$ from (\ref{gammas}).

In Eqs. (\ref{eomr}) -- (\ref{eome}) a common factor $(1-C_{(i)})^{-1}$
appears, which includes the energy derivatives of the retarded
selfenergy,
\begin{eqnarray} \label{pref} C_{(i)} & = &
\frac{1}{2 \varepsilon_i} \: \left[ \frac{\partial Re
\Sigma^{ret}_{(i)}}{\partial \epsilon_i} \: + \:
\frac{\varepsilon_i^2 - {\vec P}_i^2 - M_0^{2} - Re
\Sigma^{ret}_{(i)}}{{\tilde \Gamma}_{(i)}} \: \frac{\partial
{\tilde \Gamma}_{(i)}}{\partial \epsilon_i} \right].
\end{eqnarray}
This common factor leads to a rescaling of the 'eigentime' of
particle $i$ but does not change the trajectories as demonstrated
in Fig. 3 of Ref. \cite{Cass_off2}). According to the model
studies in \cite{Cass_off2} this prefactor is negative for
invariant masses $M$ below the pole mass $M_0$ and positive above
which implies that the factor $1/(1-C_i)$ is less than 1 for
$M<M_0$ and larger than 1 for $M>M_0$ (as stated in
\cite{Cass_off2}). This yields a time dilatation for masses $M <
M_0$ in their phase-space propagation. In order to examine the
possible effect on dilepton spectra we have repeated the model
calculations discussed in Fig. 3 of \cite{Cass_off2} including
additionally the decay of each 'testparticle' according to the
width $\tilde{\Gamma}/(2 M_i)$ in the eigen frame of the particle.
Furthermore, Eqs. (22),(23) in Section 2.5 of the present manuscript have been
adopted to calculate the influence on dilepton spectra by
integrating the differential decay rate in time. As initial
condition 'broad vector quasiparticles' close to the center of a
'nucleus'  have been assumed whereas the other parameters have
been taken the same as in Fig. 3 of \cite{Cass_off2}. The
resulting 'dilepton spectra' show a slight enhancement with
respect to the reference calculation for masses $M < M_0$ when
including the correction factor. This results from a slightly
larger 'lifetime' of the particles for $M < M_0$ - when including
the rescaling of the 'lifetime' by the correction factor - and
leads to a longer dilepton radiation time. The small differences
obtained can be traced back to the size of the correction factor
for $M < M_0$ for the model case in Ref. \cite{Cass_off2} which is
less than 12\%. Accordingly  we find the 'enhancement' in the
dilepton yield to be less than 12\% for the model scenario, too.
Since the explicit energy dependence of the retarded selfenergy is
highly model dependent we will discard the terms (\ref{pref}) in
the following study, i.e. assume $C_{(i)} = 0$.

The interpretation of the equations of motion (\ref{eomr}) --
(\ref{eome}) becomes particularly transparent if $\tilde{\Gamma}$
is independent on the 3-momentum $\vec{P}$. Then, using $M^{2} =
P^2 - Re \Sigma^{ret}$ as an independent variable instead of the
energy $P_0 \equiv \varepsilon$, Eq. (\ref{eome}) turns to
\begin{eqnarray}
\frac{dM_i^2}{dt} \; = \; \frac{M_i^2 - M_0^2}{{\tilde
\Gamma}_{(i)}} \; \frac{d {\tilde \Gamma}_{(i)}}{dt} \label{eomm}
\end{eqnarray}
for the time evolution of the test-particle $i$ in the invariant
mass squared \cite{Cass_off1,Cass_off2}. Now
 the deviation from the pole mass, i.e. $\Delta M^2 = M^2 -
M_0^2$, follows the equation
\begin{eqnarray}
{d\over dt}\Delta M^2 = {\Delta M^2\over \tilde{\Gamma}} \ {d\over
dt} \tilde{\Gamma} \ \leftrightarrow \ \frac{d}{dt}
\left(\ln\left(\frac{\Delta M^2}{\tilde{\Gamma}}\right) \right)
=0, \label{dm2}\end{eqnarray} which expresses the fact that the
off-shellness in mass scales with the total width.

\begin{figure}[t]
{\psfig{figure=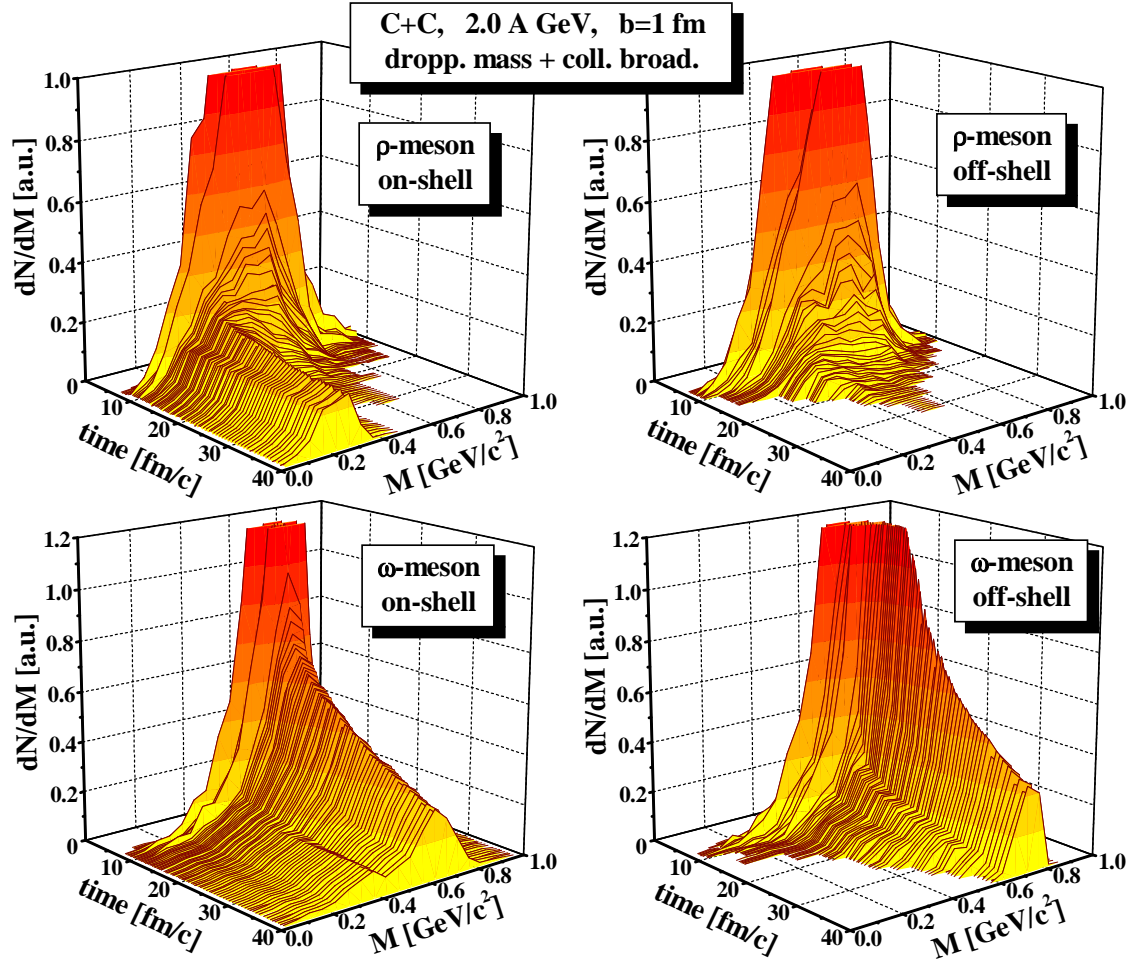,width=13cm}} \caption{Time evolution
of the mass distribution of $\rho$ (upper part) and $\omega$
(lower part) mesons for  central $C+C$ collisions (b=1 fm) at 2 A
GeV for the dropping mass + collisional broadening scenario. The
l.h.s. of Fig. \protect\ref{Fig3Donoff} correspond to the
calculations with on-shell dynamics whereas the r.h.s. show the
off-shell HSD results.} \label{Fig3Donoff} \vspace*{1mm}
\end{figure}

In order to demonstrate the importance of off-shell transport
dynamics we present in Fig. \ref{Fig3Donoff} the time evolution of
the mass distribution of $\rho$ (upper part) and $\omega$ (lower
part) mesons for  central C+C collisions (b=1 fm) at 2 A GeV for
the dropping mass + collisional broadening scenario (as an
example). The l.h.s. of Fig. \ref{Fig3Donoff} corresponds to the
calculations with on-shell propagation whereas the r.h.s. show the
results for the off-shell dynamics.

As seen from Fig. \ref{Fig3Donoff} the initial $\rho$ and $\omega$
mass distributions are quite broad even for a small system such as
$C+C$ where, however, the baryon density at 2 A GeV may reach (in
some local cells) up to $2 \rho_0$. The number of vector mesons
decreases with time due to their decays and the absorption by
baryons ($\rho N \rightarrow \pi N$ or $\rho N \rightarrow \pi\pi N$).
Most of the $\rho$ mesons decay/disappear already inside the 'fireball'
for density  $\rho_N > 0$. Due to the 'fireball' expansion the baryon
density drops quite fast, so some amount of $\rho$ mesons reach the
very low density zone or even the 'vacuum'. Since for the off-shell
case (r.h.s. of Fig. \ref{Fig3Donoff}) the $\rho$ spectral function
changes dynamically by propagation in the dense medium according to
Eqs. (14)-(16), it regains the vacuum shape for $\rho\to 0$. This does
not happen for the on-shell treatment (l.h.s. of Fig.
\ref{Fig3Donoff}); the $\rho$ spectral function does not change its
shape by propagation but only by explicit collisions with other
particles. Indeed, there is a number of $\rho$'s which survive the
decay or absorption and leave the 'fireball' with masses below
$2m_\pi$.

Accordingly, the on-shell treatment leads to the appearance of
$\rho$ mesons in the vacuum with $M\le 2m_\pi$, which can not
decay to two pions; thus they live practically 'forever' since the
probability to decay to other channels is very small. Indeed, such
$\rho$'s will continuously shine  low mass dileptons which leads
to an unphysical 'enhancement/divergence' of the dilepton yield at
low masses (note, that the dilepton yield is additionally enhanced
by a factor $\sim 1/M^3$).

The same statements are valid for the $\omega$ mesons (cf.  lower
part of Fig.  \ref{Fig3Donoff}): since the $\omega$ is a long
living resonance, a larger amount of $\omega$'s survives with an
in-medium like spectral function  in the vacuum (in case of
on-shell dynamics). Such $\omega$'s with $M < 3m_\pi$ can decay
only to $\pi \gamma$ or electromagnetically (if $M < m_\pi $).
Since such unphysical phenomena appear in on-shell transport
descriptions (including an explicit vector-meson propagation) an
off-shell treatment is mandatory.

The off-shell equations of motion (\ref{eomr}) -- (\ref{eome}) are
the present standard for the HSD transport approach and have also
been adopted by the Rossendorf group in Ref. \cite{wolfg2}.

\subsection{Vector meson production in the nuclear medium}

Apart from the explicit off-shell propagation the production of
mesons is also modified in the nuclear medium in line with their
in-medium spectral function. We mention that in-medium mass shifts
of particles have been traditionally incorporated in on-shell
transport approaches by parametrizing the vacuum cross sections as
a function of the invariant energy $\sqrt{s}$ and the threshold
$\sqrt{s_0}$ (cf. \cite{Brat97,CBRep98,sibirtsev}). In-medium
modifications of cross sections then in first order have been
incorporated by a shift of the 'pole' threshold (which corresponds
to the pole mass of the spectral function) $\sqrt{s_0^*} =
\sqrt{s_0} + M _0^* - M_0$ due to a lack of microscopic
calculations for the corresponding in-medium transition matrix
elements.

In order to account for the in-medium effects in production
cross sections, we model
the vector meson production cross sections in $NN$ and $\pi N$
reactions in the following way:
The total cross section  $\sigma_{NN \rightarrow VNN}(s,\rho_N)$
(similar $\sigma_{\pi N \rightarrow VN}(s,\rho_N)$) is
\be
\sigma_{NN \rightarrow VNN}(s,\rho_N) =\int\limits_{M_{min}}^{M_{max}}
dM\
\frac{d \sigma_{NN \rightarrow VNN}(s,M,\rho_N)}{d M} .
\label{xs_NNVVtot}
\ee

The mass differential cross sections are approximated by
\be
\hspace{-0.7cm} \frac{d \sigma_{NN \rightarrow VNN}(s,M,\rho_N)}{d
M} \: = \: \sigma_{NN \rightarrow VNN}^{0}(s,M,\rho_N) \cdot
A(M,\rho_N)\cdot {\int\limits_{M_{min}}^{M_{max}} A(M,\rho_N) dM
\over \int\limits_{M_{min}}^{M_{lim}} A(M,\rho_N) dM }
\label{xs_NNVV} \ee where $A(M,\rho_N)$ denotes the meson spectral
function (\ref{spfunV}) for given total width $\Gamma_V^*$
(\ref{gammas}); $M_{max}=\sqrt{s}-2 m_N$ is the maximal
kinematically allowed invariant mass of the vector meson $V$ and
$m_N$ is the mass of the nucleon. In Eq. (\ref{xs_NNVV})
$\sigma_{NN \rightarrow VNN}^{0}(s,M,\rho_N)$ is the
parametrization (based on phase space) of the vacuum cross section
\cite{CBRep98} with a modified 'pole' threshold
$\sqrt{s_0^*}(\rho_N) = 2 m_N + M_0^*(\rho_N)$ instead of
$\sqrt{s_0} = 2 m_N + M_0$  in the vacuum. Note, that  the
physical threshold for the $\rho$ production, e.g., in $NN$
reactions is defined as $\sqrt s_{th} = 2 m_N + M_{min}$, i.e. for
the broadening scenario $\sqrt s_{th} \to 2 m_N$. Thus, the
formula (\ref{xs_NNVV}) is used to model the vector meson
production in $NN$ reactions in the vacuum and in the medium.

\begin{figure}[t]
\psfig{figure=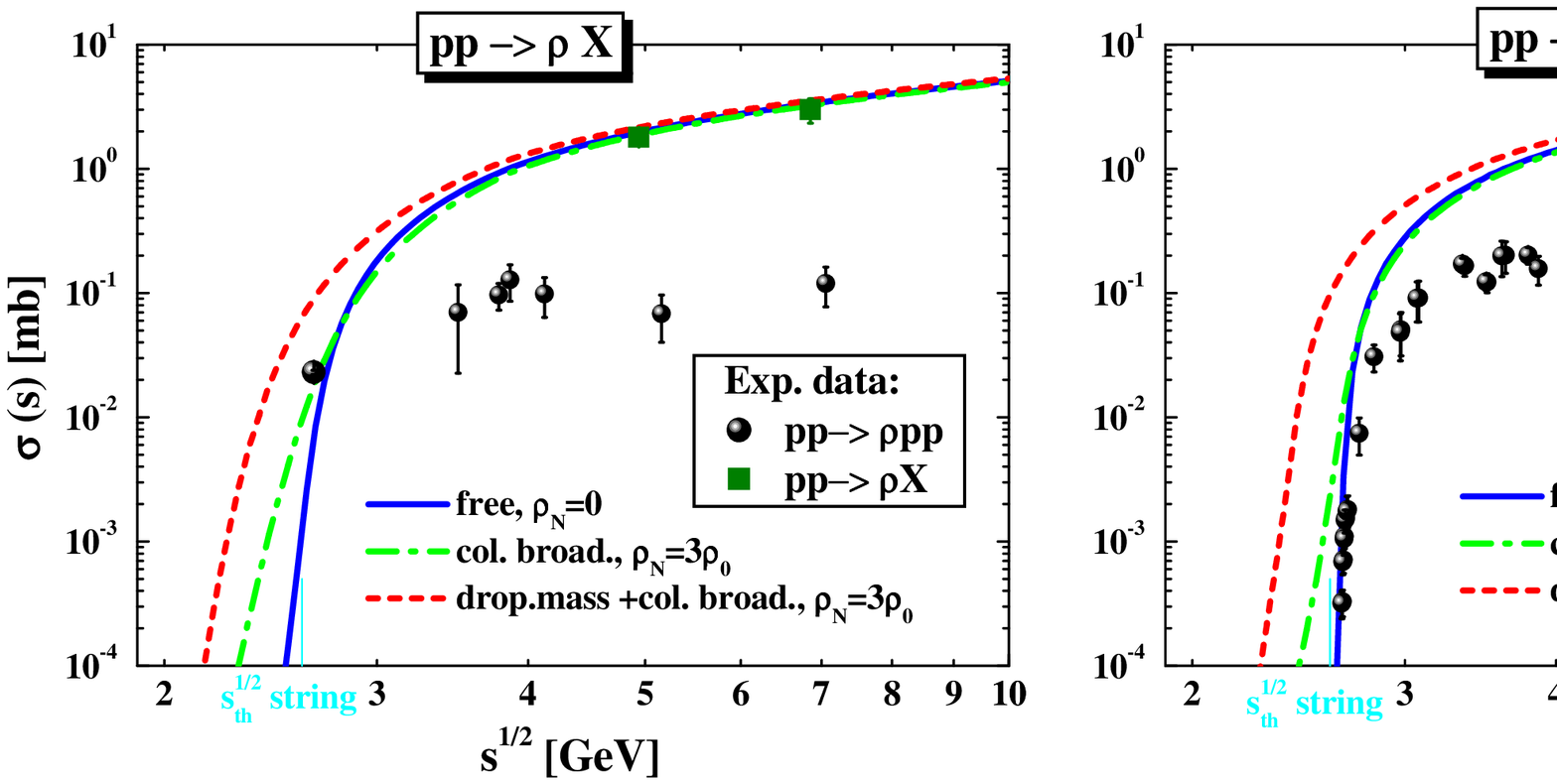,width=11cm}
\caption{The production cross sections for the channels $N N \rightarrow V X$
 as a function of the invariant energy $\sqrt{s}$ for densities
$\rho_N = 0$ and $3\rho_0$. The solid lines represent the parametrizations
of inclusive $N N \rightarrow V X$ cross sections
in free space while the dash-dotted lines stand for the 'collisional
broadening' scenario; the dashed lines show the inclusive cross
sections in the 'dropping mass + collisional broadening' scenario.
 The vertical thin blue line displays the
threshold for meson production by string formation and decay
($\sqrt{s}_{th}=2.6$ GeV) in case of  baryon-baryon channels.
The experimental data  \protect\cite{expVprod,DISTO02,Moskal02} are shown for
exclusive $pp \rightarrow V pp$ (dots) and inclusive $pp \rightarrow V X$
(squares) vector meson production where $X$ stands for two baryons and further mesons. }
\label{Fig7}
\end{figure}

\begin{figure}[t]
\psfig{figure=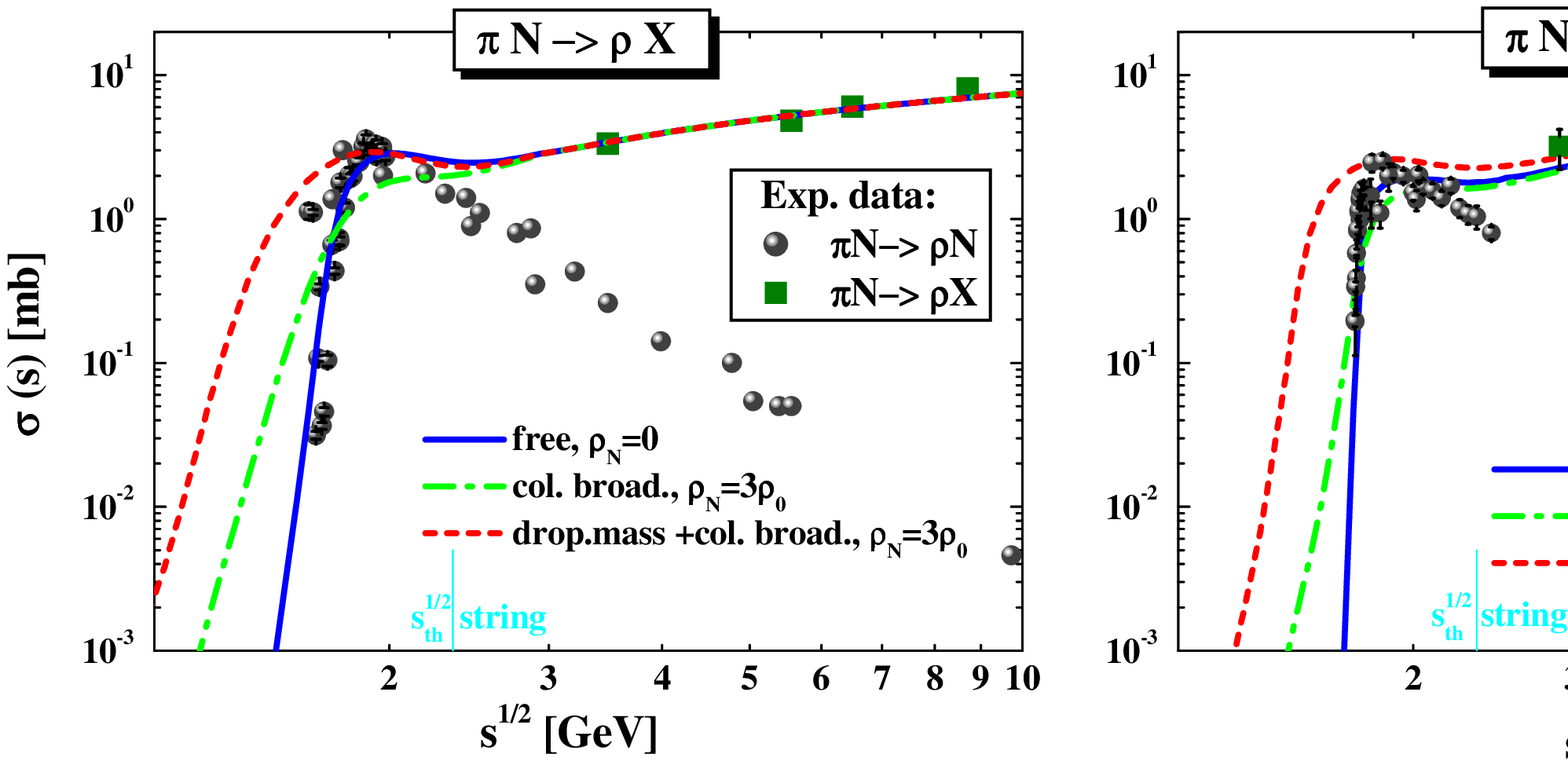,width=11cm}
\caption{The production cross sections for the channels $\pi N
\rightarrow V N$ ($V= \rho, \omega $)  as a function of the
invariant energy $\sqrt{s}$ for densities $\rho_N = 0$ and $3
\rho_0$. The solid lines represent the parametrizations
of  inclusive $\pi N \rightarrow V X$ cross sections in
free space while the dash-dotted lines stand for the 'collisional
broadening' scenario. The dashed lines show the inclusive cross
sections in the 'dropping mass + collisional broadening' scenario.
The experimental data  \protect\cite{expVprod,Moskal02} are shown for
exclusive $\pi N \rightarrow V N$ (dots) and inclusive
$\pi N \rightarrow V X$ (squares) vector  meson production.
The vertical light blue line shows the threshold for meson production
by string formation and decay ($\sqrt{s}_{th}=2.3$ GeV) as
implemented in HSD for meson-baryon channels.}
\label{Fig5}
\end{figure}

The results for the channels $N N \rightarrow V X$ are displayed
in Fig. \ref{Fig7} as a function of the invariant energy
$\sqrt{s}$ for densities $\rho_N = 0$ and $\rho_N =3 \rho_0$. The
solid lines represent the parametrizations for inclusive
 ($N N \rightarrow V X$) cross sections in free space while the
dash-dotted lines stand for the 'collisional broadening' scenario;
the dashed lines show the inclusive cross sections in the
'dropping mass + collisional broadening' scenario.
 The light vertical blue line stands for the
threshold for meson production by string formation and decay
($\sqrt{s}_{th}=2.6$ GeV) for baryon-baryon channels.
The squares show the experimental data
\cite{expVprod,DISTO02,Moskal02} for the inclusive $N N \rightarrow V X$
vector meson production (here $X$ stays for all final particles including two baryons and further
mesons).
The dots correspond to the experimental data for the exclusive $pp
\rightarrow V pp$ reaction.

The string formation and decay reasonable matches the inclusive ($N N
\rightarrow V X$) data for $\rho$ (squares)  taken from Refs.
\cite{expVprod,DISTO02,Moskal02}. It is seen that the excitation
function of the vector meson production cross section in the
'collisional broadening' scenario is basically smeared out close to
threshold.  Only when incorporating additionally a dropping mass the
thresholds are shifted down in energy such that the production cross
sections become enhanced in the subthreshold regime with increasing
nuclear density $\rho_N$.

The actual results for the channels $\pi N \rightarrow V N$ ($V=
\rho, \omega$) are displayed in Fig. \ref{Fig5} as a function of
the invariant energy $\sqrt{s}$ for densities $\rho_N = 0$ and
$\rho_N =3 \rho_0$. Here the solid lines represent the
parametrizations of  $\pi N \rightarrow V X$ cross sections
in free space while the dash-dotted lines stand
for the 'collisional broadening' scenario. The dashed lines show
the inclusive cross sections in the 'dropping mass + collisional
broadening' scenario. The experimental data are displayed for
exclusive $\pi N \rightarrow V N$ (dots) and inclusive
$\pi N \rightarrow V X$ (squares) vector  meson production
channels. The vertical light blue line stands for the threshold
for meson production by string formation and decay
($\sqrt{s}_{th}=2.3$ GeV) as implemented in HSD for meson-baryon
channels. It is seen that the string decay reasonable matches the
inclusive ($\pi N \rightarrow V X$) data for $\rho$ and $\omega$
(squares) taken from Refs.
\cite{expVprod,Moskal02}.

One observes from Fig. \ref{Fig5} that in the 'collisional
broadening' scenario the excitation function of the vector meson
production cross section is only smeared out close to threshold
due to the broader vector meson width at finite density. As in
case of $NN$ reactions   the thresholds are shifted down in energy
when incorporating additionally a dropping mass. Since the
collisional broadening is low for the $\phi$ meson the
modifications of the production cross section for $\phi$'s stay
very moderate (not shown explicitly).

A note of caution has to be added here because the present
parametrizations for vector meson production in the nuclear medium
are not  controlled by microscopic studies on the various
production channels for the case of finite nuclear density
$\rho_N$. A consistent microscopic description of in-medium vector
meson production cross sections thus is urgently needed.
Furthermore, it should be noted that the experimental
determination of a $\rho$ meson production cross section is
problematic since this would imply to measure, e.g., the full
invariant mass range of $\pi^+ \pi^-$ pairs in a $p$-wave final
state. This is a rather complicated task because the two-pion
decay width $\Gamma_{\pi\pi}$ approaches zero close to the two-pion
threshold where the data are dominated by pions in a relative
$s$-wave. Accordingly experimental data have to be considered
simultaneously with the operational definition for a
'$\rho$-meson'.

We also mention that in the early study \cite{BratKo99} we have
considered the possible contribution from the Dalitz decay of the
$N(1520)$ resonance: $N(1520) \to  \rho N \to e^+e^- N$. At those
times this process has been treated perturbatively, i.e. without
explicit propagation of the $N(1520)$. We obtained  from our
qualitative estimates  that the dilepton radiation from the
$N(1520)$ in the hot and dense medium is strongly reduced due to a
'melting' of the $N(1520)$. Since the width and pole position of
the $N(1520)$ is still quite uncertain we do not include the
contribution of the $N(1520)$ explicitly and rather rely on the
findings in \cite{BratKo99} that the $N(1520)$ decay contribution
to dileptons should become small in a dense medium.

\subsection{'In-medium' extension of the LUND string model}

The particle production from baryon-baryon collisions above
$\sqrt{s}_{th}\simeq 2.6$ GeV and from meson-baryon collisions
above $\sqrt{s}_{th}\simeq 2.3$ GeV are modeled in HSD via  string
formation and decay. The actual realization is the LUND
string-fragmentation model \Fritiof{} \cite{FRITIOF} which is
known to describe inelastic hadronic reactions in a wide energy
regime at {\it zero baryon density}. In the standard version of
\Fritiof{} the resonances are produced according to the vacuum
Breit-Wigner spectral function (by default non-relativistic) with
a constant width. Moreover, the Breit-Wigner shape is truncated
symmetrically around the pole mass, $|M-M_0|<\delta$, with
$\delta$  chosen 'properly' for each particle such that no
problems are encountered in the particle decay chains \cite{FRITIOF}.

However, in order to study in-medium effects at  energies close to
the individual thresholds $\sqrt{s}_{th}$ one has to incorporate
the relativistic in-medium spectral functions (as described in
Section 2.1 for the vector mesons) also consistently in the string
fragmentation model.  Such an extension of the \Fritiof{} model is
performed in this study for the first time and the resonance
production in the medium via \Fritiof{} is treated with the full
relativistic Breit-Wigner spectral functions including the density
dependent self-energy and in-medium width (depending on mass and
baryon density). Also the truncation of the spectral function in
mass is removed, i.e. the resonance mass is chosen within the
physical thresholds from $M_{min}$ to $M_{max}$. As before the total
energy and momentum conservation holds strictly in the extended
\Fritiof{} model.

We mention that this extension of the Lund model - which allows
to implement any shapes of spectral functions into the
fragmentation scheme - has been used not only for vector mesons
but also for the $\Delta$ resonance in order to incorporate the
mass-dependent width $\Gamma_{\Delta}(M_{\Delta})$ (in line with
the low energy part of HSD, where the parametrization from Ref.
\cite{Moniz84} has been used for $\Gamma_{\Delta}(M_{\Delta})$), which
leads to a hardening of the high mass dilepton spectra from the
$\Delta$ Dalitz decay. Note, that 'in-medium' effects - i.e.
collisional broadening or/and dropping masses - are accounted only for
the vector mesons ($\rho, \omega, \phi$) in the present study.

\subsection{Time integration (or 'shining') method for dilepton production}

Since dilepton production is a very rare process (e.g.  the
branching ratio for the vector meson decay is $\sim 10^{-5}$), a
perturbative method is used in order to increase statistics. In
the HSD approach (in this study as well as in our earlier
investigations
\cite{CBRep98,BratRapp98,BratKo99,Brat_pA01,Brat_KEK02}) we use
the time integration  (or 'shining') method first introduced by
Li and Ko \cite{LiKo95}. The main idea of this method is that
dileptons can be emitted during the full lifetime of the resonance
$R$ before its strong decay into hadrons or absorption by the
surrounding medium. For example, the $\rho^0$ decay (with
invariant mass $M$) to $e^+ e^-$ during the propagation through
the medium from the production time $t=0$ up to the final
('death') time $t_F$ - which might correspond to an absorption by
baryons or to reactions with other hadrons as well as the strong
decay into two pions - is calculated as
\begin{equation}
{d N^{\rho\to e^+ e^-}\over d M} = \sum_{t=0}^{t_F}
\Gamma^{\rho^0 \to e^+ e^-} (M) \cdot {\Delta t \over \gamma
(\hbar c)} \cdot {1\over \Delta M} \label{Nrho}
\end{equation}
in the mass bin $\Delta M$ and time step $\Delta t$ (in fm$/c$).
In (\ref{Nrho}) $\gamma$ is the Lorentz factor of the $\rho$-meson
with respect to the calculational frame. The electromagnetic decay
width is defined as
\begin{equation}
\Gamma^{\rho^0 \to e^+ e^-}(M) = C_\rho {{M_0^*}^4 \over M^3},
\label{gamrn}
\end{equation}
where $C_\rho= {\Gamma^{\rho\to e^+e^-}(M_0) / M_0}$. Here $M_0$
is the vacuum pole mass, $M_0^*$ is the in-medium pole mass which
is equal to the vacuum pole mass for the collisional broadening
scenario, however, is shifted for the dropping mass scenario
according to Eq. (\ref{Brown}).

The time integration method allows to account for the full
in-medium dynamics of vector mesons from production ('birth') up to their
'death'. We note that by calculating the dilepton emission
only at the strong decay vertex (e.g. as in Refs.
\cite{bleicher2,Vogel07}) the dilepton rate (as well as the
density dependence of the dilepton emission) is underestimated
since a sizeable part of the emission history (e.g. before the
absorption point by baryons or before the decay to pions) is lost.

\subsection{$e^+e^-$ bremsstrahlung in elementary reactions}

The implementation of  $e^+e^-$ bremsstrahlung from hadronic
reactions in transport approaches has been limited so far to the
dominant $pn$ channel in the 'soft photon' approximation (with phase-space
corrections) \cite{CBRep98,Ernst,Xiong90,Wolf90,Gudima} or even
been neglected completely \cite{bleicher2,fuchs3}. As we will demonstrate
below this contribution should be reexamined and revised
accordingly.

The soft-photon approximation (SPA) \cite{GaleK87} is based on the
assumption that the radiation from internal lines is negligible and the
strong interaction vertex is on-shell. In this case the strong
interaction part and the electromagnetic part can be separated,
however, the cross section for dileptons has to be corrected
\cite{GaleK87} by reducing the phase-space for the colliding particles
in their final-state.  The phase-space corrected soft-photon cross
section for the reaction $1+2\to  1 + 2 + e^+ +e^-$ can
be written as
\begin{eqnarray}
&& \frac{d \sigma}{dy d^2 q_T d M} = \frac{\alpha^2}{6 \pi^2} \frac{{\bar
\sigma (s)}}{M q_0^2} \frac{R_2(s_2)}{R_2(s)}, \label{brems} \\
&& R_2(s) = \sqrt{1 - (m_1 + m_2)^2/s}, \nonumber \\
&& s_2 = s + M^2 - 2 q_0 \sqrt{s}, \nonumber \\
&& {\bar \sigma(s)} = \frac{s - (m_1 + m_2)^2}{2 m_1^2} \sigma(s), \nonumber
\end{eqnarray}
where $m_1$ is the mass of the charged accelerated particle, $M$
is the dilepton invariant mass, $q_0$ the energy, $q_T$ the
transverse momentum and $y$ the rapidity of the dilepton pair. In
(\ref{brems})  $\sigma(s)$ is the on-shell elastic cross section
for the reaction $1+2\to 1+2$.

\begin{figure}[h]
\phantom{a}\hspace*{6mm} {\psfig{figure=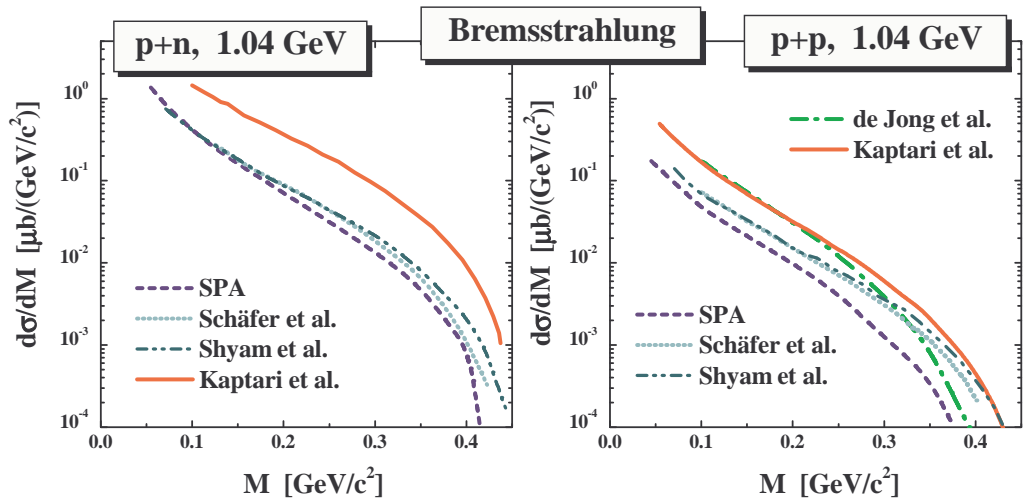,width=10.5cm}}
\caption{The $e^+e^-$ bremstrahlung from $pn$ (l.h.s.) and $pp$
(r.h.s.) channels. The dashed lines show the SPA results
\cite{GaleK87}, the dotted and dashed-dot-dotted lines correspond
to the OBE calculations by Sch\"afer et al. \cite{Schaefer89} and
Shyam et al. \cite{Shyam03}, respectively. The dash-dotted line
(r.h.s.) displays the $T$-matrix result  from de Jong et al.
\cite{deJong97} while the red solid lines show the calculations by
Kaptari et al. \cite{Kaptari}. } \label{Fig_Brems}
\end{figure}

In spite of the general limitation of the SPA -- discussed in
detail in Refs. \cite{GaleK87,Lichard95} -- Eq.~(\ref{brems}) has
been widely used for the calculation of the bremsstrahlung
dilepton spectra by different transport groups
\cite{CBRep98,Ernst,Xiong90,Wolf90,Gudima}. Note, at those early
times the applicability of the SPA for an estimate of dilepton
radiation from $NN$ collisions at 1-2 GeV bombarding energies has
been supported by independent One-Boson-Exchange (OBE) model
calculations by Sch\"afer et al.  \cite{Schaefer89} and later on
by Shyam et al. \cite{Shyam03} (cf. Fig.  \ref{Fig_Brems}). In
these models the effective parameters have been adjusted to
describe elastic $NN$ scattering at intermediate energies. The
models have then been applied to  bremsstrahlung processes
including the interference of different diagrams for the dilepton
emission from all charged hadrons.  As shown in Ref.
\cite{Schaefer89} the $pp$ bremsstrahlung is much smaller than
$pn$ bremsstrahlung due to a destructive interference of
amplitudes from the initial and final radiation. We note that
'gauge invariant' results have been obtained in Refs.
\cite{Schaefer89,Shyam03} by 'gauging' the phenomenological form
factors at the meson-baryon vertices. However, there are different
schemes to introduce 'gauge invariance' in OBE models - as
stressed by Kondratyuk and Scholten \cite{Scholten} - which lead
to sizeably different cross sections.

One has to point out that already in 1997 an independent study by
de Jong et al. \cite{deJong97} - based on a full $T$-matrix
approach - has indicated that the validity of the SPA for $e^+e^-$
bremsstrahlung at intermediate energies of 1-2 GeV may be very
questionable. However, in  Ref. \cite{deJong97} only the $pp$
reaction has been considered (cf. r.h.s. of Fig. \ref{Fig_Brems});
indeed, the bremsstrahlung in the full T-matrix approach is larger
by a factor of about 3 than the corresponding SPA calculations
(and OBE results).

Recently,  new covariant OBE calculations for dilepton
bremsstrahlung have been performed by Kaptari and K\"ampfer
\cite{Kaptari}. The effective parameters for $NN$ scattering have
been taken similar to \cite{Shyam03}, however, the restoration of
gauge invariance has been realized in a different way. As
mentioned above, there are several prescriptions for restoring
gauge invariance in effective theories including
momentum-dependent form factors in interactions with charged
hadrons \cite{Scholten} and the actual results very strongly
depend on the prescription employed. The scheme in Ref.
\cite{Kaptari} is to include explicitly the vertex form factors
into the Ward-Takahashi identity for the full meson-exchange
propagators, which is different from the method used in Refs.
\cite{Schaefer89,Shyam03}.

The $pn$ and $pp$ bremsstrahlung differential spectra from Kaptari
{\it et al.} \cite{Kaptari} are shown in Fig. \ref{Fig_Brems} by
the solid red lines. One can see that the difference between the
Kaptari model and the  SPA and OBE \cite{Schaefer89,Shyam03}
results is up to a factor of 4 for $pn$ reactions  and about a
factor of 3 for $pp$ collisions at 1 GeV. Note, that the $pp$
bremsstrahlung from \cite{Kaptari} is on the same level as the
$T$-matrix calculations from Ref. \cite{deJong97} (cf. Fig.
\ref{Fig_Brems}, r.h.s.).
Thus, according to the Kaptari model the contributions from $pn$ and
$pp$ bremsstrahlung have been substantially {\it underestimated} in the
previous transport model calculations \cite{CBRep98,Ernst} (accounting
only for elastic scattering events with charged hadrons).

We stress that formula (\ref{brems}) is only applicable for the
bremsstrahlung radiation from elementary elastic scattering
reactions (i.e. elastic $NN$ or $\pi N$ collisions). However, in
the transport models of Refs.  \cite{wolfg1,wolfg2} it has been
adopted also for inelastic $NN$ reactions which faces the problem
of double counting since the dilepton emission from inelastic $NN$
reactions is dominantly taken into account by decays of the produced
particles (i.e. $\Delta,\eta, \omega$ Dalitz decays, direct
$\rho,\omega,\phi$ decay etc.). On the other hand, electromagnetic
radiation also emerges from inelastic channels since each charged
particle (that is decelerated or accelerated in a collision) will
radiate dileptons. Due to an interference of all amplitudes there
is a partial cancellation between initial state amplitudes and
final state amplitudes in case of equal charges due to a different
sign in the acceleration. On the other hand two hadrons with
opposite charges in the final state will also interfere
destructively such that general estimates for a wide class of
inelastic hadronic reactions cannot easily be performed. In this
respect the strategy in Refs. \cite{wolfg1,wolfg2} (to account for
radiation from inelastic channels) is a very crude assumption
whereas our approximation implies to disregard such additional
radiation amplitudes in case of dominant electromagnetic decays
which are taken into account explicitly. Due to a lack of
microscopic calculations on the $T$-matrix level for radiation from
channels with further charged particles in the final state we have to wait for
respective experimental results to verify/falsify the different
approximations.

In the present study we will adopt the results from Ref.
\cite{Kaptari} for the elementary bremsstrahlung processes. As
will be shown in the next Sections this will have a sizeable
impact on our results for the dilepton radiation from $AA$
collisions at 1 -- 2 A$\cdot$GeV.

We have to point out, furthermore, that in order to separate the
bremsstrahlung ($pp\to ppe^+e^-$) from a vector-dominance like
dilepton production via the $\rho$-meson ($pp\to pp\rho, \rho\to
e^+e^-$), we do not employ a vector-dominance formfactor when
calculating the bremsstrahlung. Thus, the dilepton radiation via
the decay of the virtual photon ($pp\to pp\gamma^*, \gamma^*\to
e^+e^-$) and the direct $\rho$ decay to $e^+e^-$ are distinguished
explicitly in the calculations. This is different from the
calculations of the T\"ubingen group \cite{Fuchs03} which have no
dynamical $\rho$ meson but only a medium-dependent formfactor in
the bremsstrahlung channel.

In closing this Subsection we stress that new calculations of
dilepton bremsstrahlung - based on a consistent full $G-$ (or
$T-$) matrix approach - are urgently needed in order to clarify
the unsatisfactory situation. From the experimental side high
statistics measurements for multi-differential
dilepton spectra from elementary $pp$
and $pn (pd)$ reactions are required to settle this open question
for elastic as well as inelastic channels.

\section{$\pi^0$ and $\eta$ production}

Since the low mass dilepton spectra are dominated by the Dalitz
decays of $\pi^0$ and $\eta$ an explicit control of  their
production cross section is mandatory for a proper interpretation
of dilepton mass spectra. We recall that the early suggestion of
the DLS Collaboration that the enhanced $\e^+e^-$ production seen
at the BEVELAC in $C+C$ and $Ca+Ca$ collisions might be due to an
enhanced production of $\eta$'s was rejected by the TAPS
Collaboration measuring the $\eta$ cross section via the $\pi^0
\gamma$ decay channel for the same systems \cite{Holzmann}.
The TAPS argument has been confirmed by the transport calculations
in Refs. \cite{BratRapp98,Brat98mt}.

\begin{figure}[t]
\phantom{a}\vspace*{5mm}
\centerline{\psfig{figure=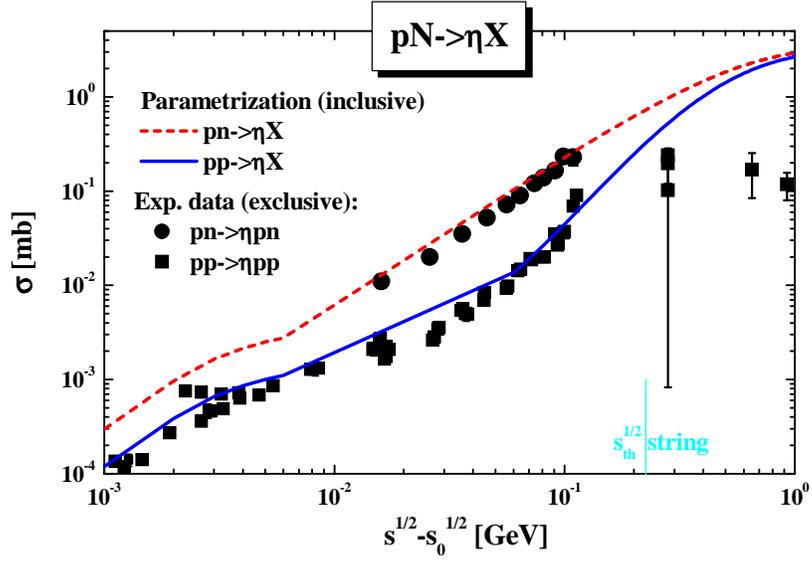,width=8.5cm}}
\caption{The $\eta$ production cross section from $pp$ and $pn$
reactions as a function of the invariant energy above threshold.
The solid  and dashed  lines represent the corresponding inclusive
($pN \to\eta X$)
parametrizations/results from the HSD transport approach, whereas
the full symbols show the exclusive ($pN\to\eta pN$)
experimental data from Refs. \cite{etanew,Moskal02,Moskal05}.
} \label{Fig1}
\end{figure}

In the last decade the experimental information of $\eta$
production in $pp$ and $pn$ reactions - especially close to
threshold - has been drastically improved such that more
appropriate parametrizations of these cross sections have to be
implemented in transport in order to stay in line with the
experimental progress. The presently available experimental information from
exclusive $pp$ and $pn$ reactions is displayed in Fig. \ref{Fig1} by the
full symbols as a function of the invariant energy above threshold
and indicates that the $pn$ cross section is larger than the $pp$
cross section for $\eta$ production by up to a factor of 6. Only
at sufficiently high energy the initial isospin no longer plays
any role due to multiple associated pion production. The solid and
dashed lines in Fig. \ref{Fig1} show the novel parametrizations/results
from HSD for $pp$ and $pn$ reactions which smoothly join the inclusive
$pN \to\eta X$ production from string decay at high energy and are
compatible with the low energy exclusive $pN \to\eta pN$
data from Refs. \cite{etanew,Moskal02,Moskal05}.

\begin{figure}[t]
\phantom{a}\hspace*{5mm} \psfig{figure=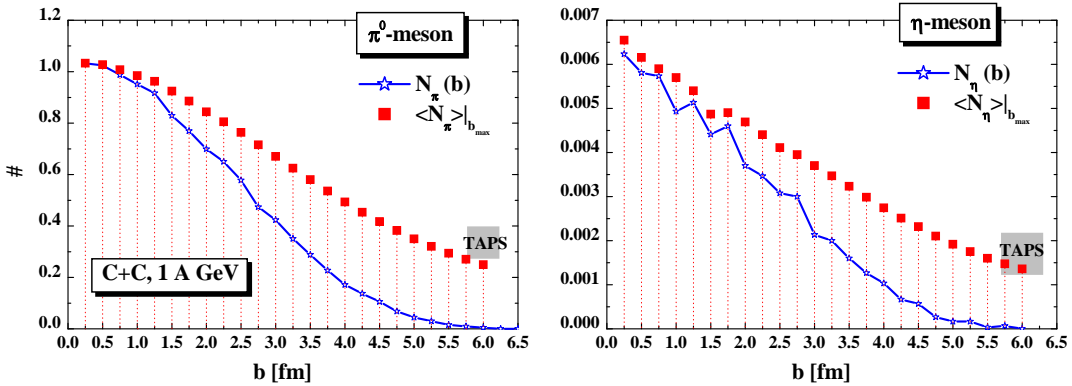,width=10.5cm}
\caption{The number of $\pi^0$'s (l.h.s.) and $\eta$'s (r.h.s.) from HSD for
$^{12}C+^{12}C$ at 1 A$\cdot$GeV as a function of the impact parameter $b$ (lower blue
lines). The red squares denote the average number of $\pi$'s and
$\eta$'s according to Eq. (\ref{average}) for $b_{max} = b$ and
show that the average number of $\pi^0$'s and $\eta$'s from the
TAPS Collaboration in inclusive $^{12}C+^{12}C$ reactions (shaded
areas) is approximately reproduced for $b_{max} \approx$ 6 fm.}
\label{Fig2}
\end{figure}

\begin{figure}[t]
\phantom{a}\hspace*{5mm} \psfig{figure=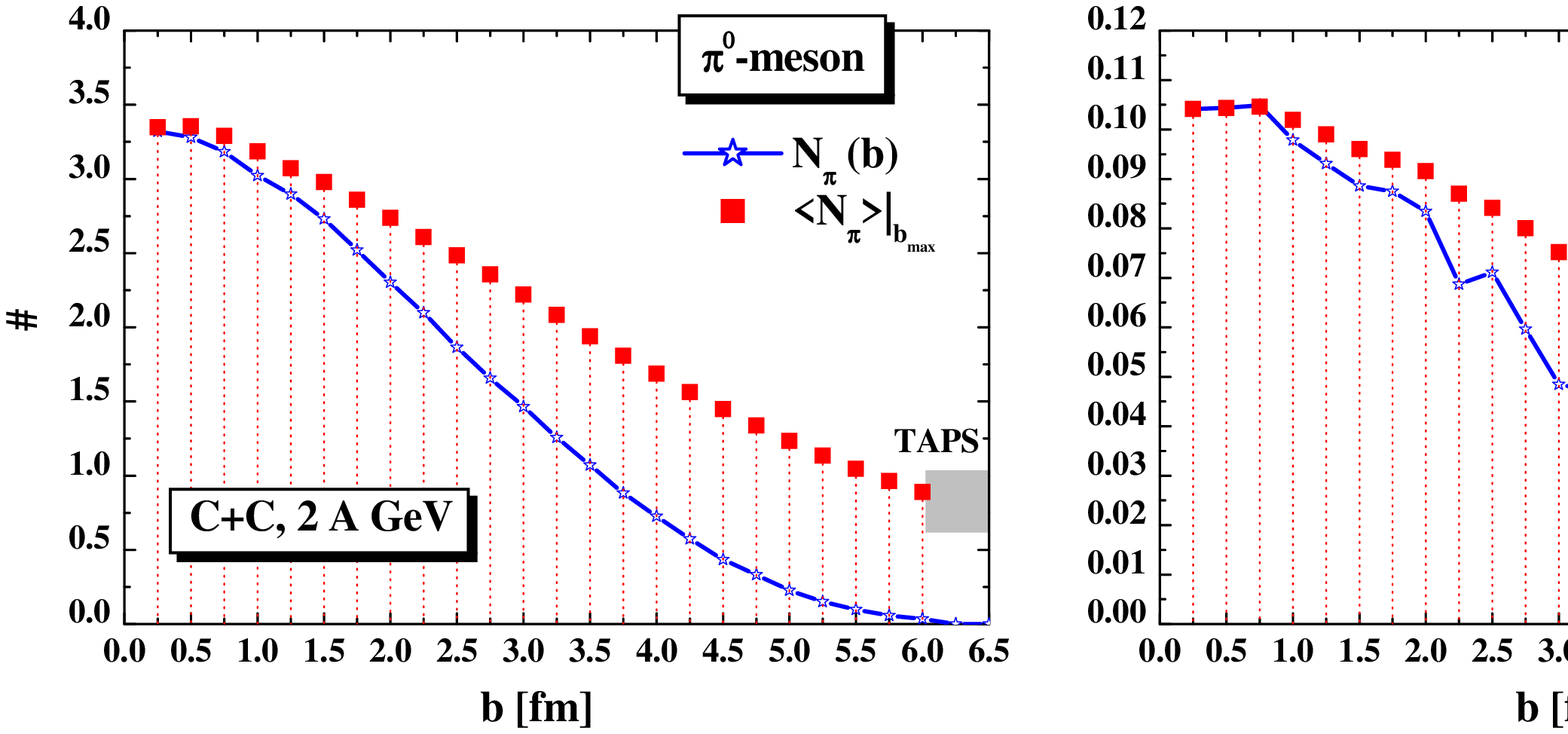,width=10.5cm}
\caption{ The same as Fig. \ref{Fig2} for $^{12}C+^{12}C$
reactions at 2 A$\cdot$GeV.} \label{Fig3}
\end{figure}

The number of $\pi^0$'s and $\eta$'s from HSD for $^{12}C+^{12}C$
at 1 A$\cdot$GeV  and 2 A$\cdot$GeV are shown in Figs. \ref{Fig2}
and \ref{Fig3}  as a function of the impact parameter $b$ (lower
blue lines). The average number of $\pi$'s and $\eta$'s is given
by, \be \label{average} <N_x>|_{b_{max}} = \frac{2\pi
\int_0^{b_{max}} N_x(b) \ b \ db}{\pi b_{max}^2}, \ee where
$b_{max}$ denotes the maximum impact parameter considered in the
calculation and $x= (\pi^0, \eta)$. The corresponding results for
these quantities are displayed in Figs. \ref{Fig2} and \ref{Fig3}
 by the red squares as a function of $b=b_{max}$ and demonstrate
that the average number of $\pi^0$'s and $\eta$'s from the TAPS
Collaboration in inclusive $^{12}C+^{12}C$ reactions (shaded
areas) are reasonably reproduced by the transport calculations.
Note that the inclusive reaction cross section is given by
$\sigma_{incl} = \pi b_{max}^2$.

\begin{figure}[t]
\phantom{a}\vspace*{5mm}
\centerline{\psfig{figure=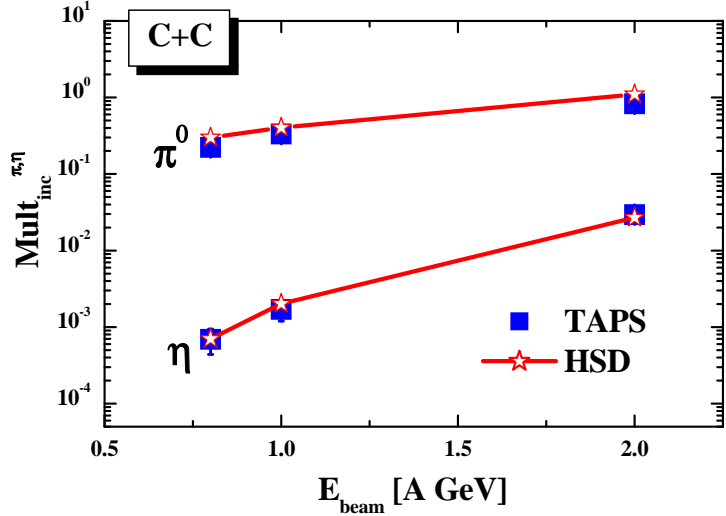,width=9.5cm}}
\caption{The average $\pi^0$ and $\eta$ multiplicities for
$^{12}C+^{12}C$ from 0.8 to 2 A$\cdot$GeV from the TAPS
Collaboration \cite{TAPS97} (full squares) in comparison to the
corresponding HSD results (stars).} \label{Fig4}
\end{figure}

In line with Figs. \ref{Fig2} and \ref{Fig3}  we show in Fig.
\ref{Fig4} the average $\pi^0$ and $\eta$ multiplicities for
$^{12}C+^{12}C$ from 0.8 to 2 A$\cdot$GeV from the TAPS
Collaboration \cite{TAPS97} (full squares) in comparison to the
corresponding HSD results (stars). Since the agreement is
acceptable (within error bars) we may proceed with the actual
dilepton studies.

\section{Dilepton production in $pp$ and $pd$ reactions}

\begin{figure}[t]
\phantom{a}\vspace*{5mm}
\centerline{\psfig{figure=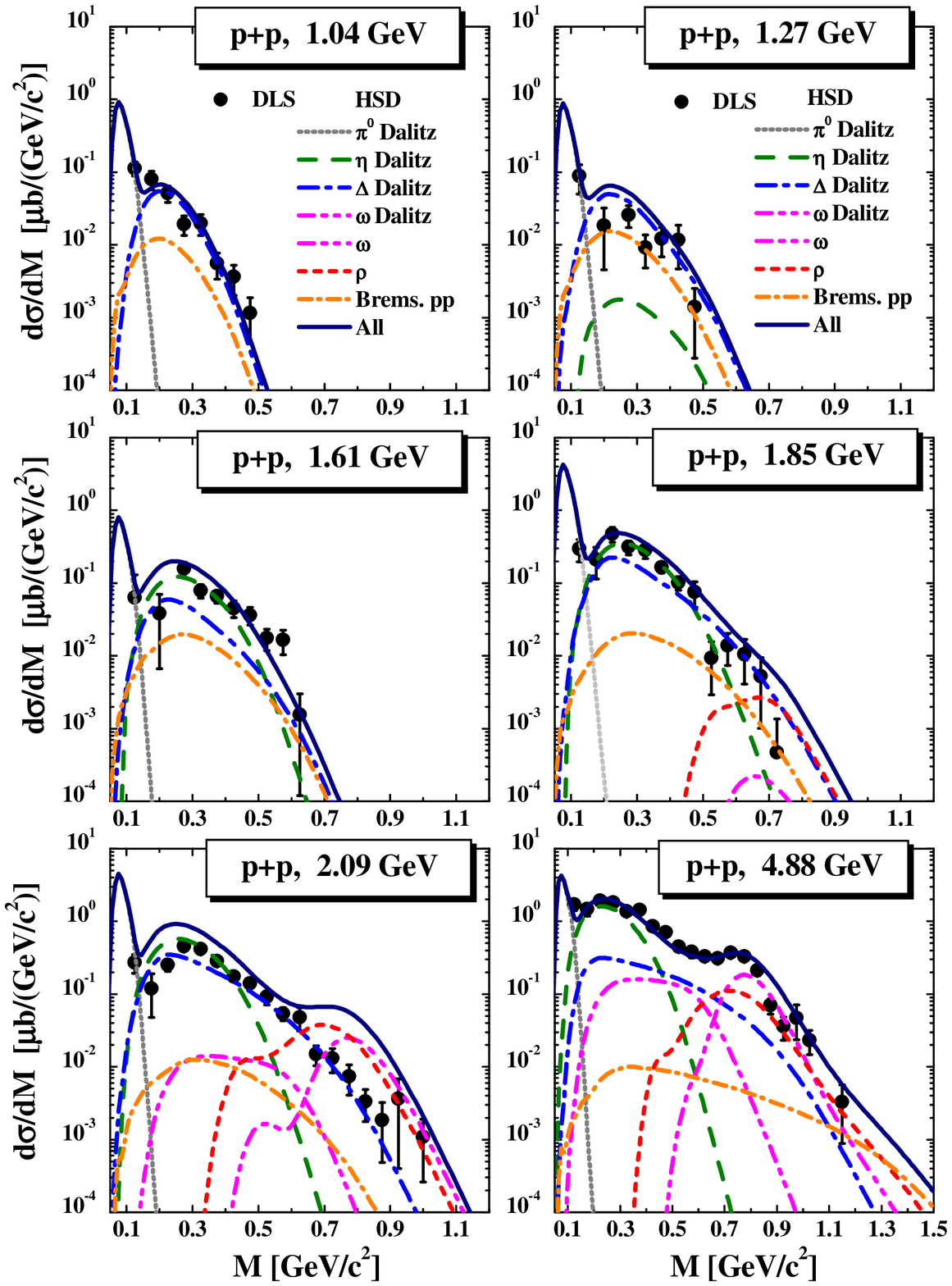,width=10.5cm}}
\caption{The differential cross section for $e^+e^-$ production in
$pp$ reactions  at bombarding energies of 1.04, 1.27, 1.61, 1.85,
2.09 and 4.88 GeV in comparison to the data from the BEVALAC
\cite{DLSpp} including the DLS acceptance filter and mass resolution
\cite{DLSpp}. The various contributions from the different channels
(in the HSD calculations) are indicated additionally (see color
coding in the legend).  } \label{Fig9}
\end{figure}

Before coming to explicit results for nucleus-nucleus collisions
we consider dilepton production in more elementary reactions like
$pp$ and $pd$ in the energy range of interest. The results for
$e^+e^-$ production in $pp$ reactions are shown in Fig. \ref{Fig9}
at bombarding energies of 1.04, 1.27, 1.61, 1.85, 2.09 and 4.88 GeV
in comparison to the data from the BEVALAC \cite{DLSpp} including the
DLS acceptance filter and mass resolution \cite{DLSpp}. The various
contributions from the different channels (in the HSD
calculations) are indicated additionally (see color coding in the
figure). At all energies the $\pi^0$ Dalitz decay dominates for
invariant masses up to the pion mass whereas the contributions
from 0.2 to 0.6 GeV vary drastically with energy. At 1.09 and 1.27
GeV here the $\Delta$ Dalitz decay dominates which is superseeded
by the $\eta$ Dalitz decay already at 1.61 GeV. The $pp$
bremsstrahlung contribution also gives a sizeable yield to the
total dilepton cross section at the lower energies. At 2.09 GeV we
find that the calculations overestimate the yield from the $\rho$
and $\omega$ decays which has to be kept in mind when comparing to
nucleus-nucleus data at 2 A$\cdot$GeV (see below). At 4.88 GeV all
channels result from string excitation and decay (except
bremsstrahlung) which is sufficiently well in line with the
BEVALAC data.
We note that at all energies the contribution from the direct Dalitz 
decay of heavy resonances $R\to e^+e^-N$ is suppressed compare to 
$\Delta\to e^+e^- N$ as pointed out before in Ref. \cite{Wolf03}.

\begin{figure}[t]
\phantom{a}\vspace*{5mm}
\centerline{\psfig{figure=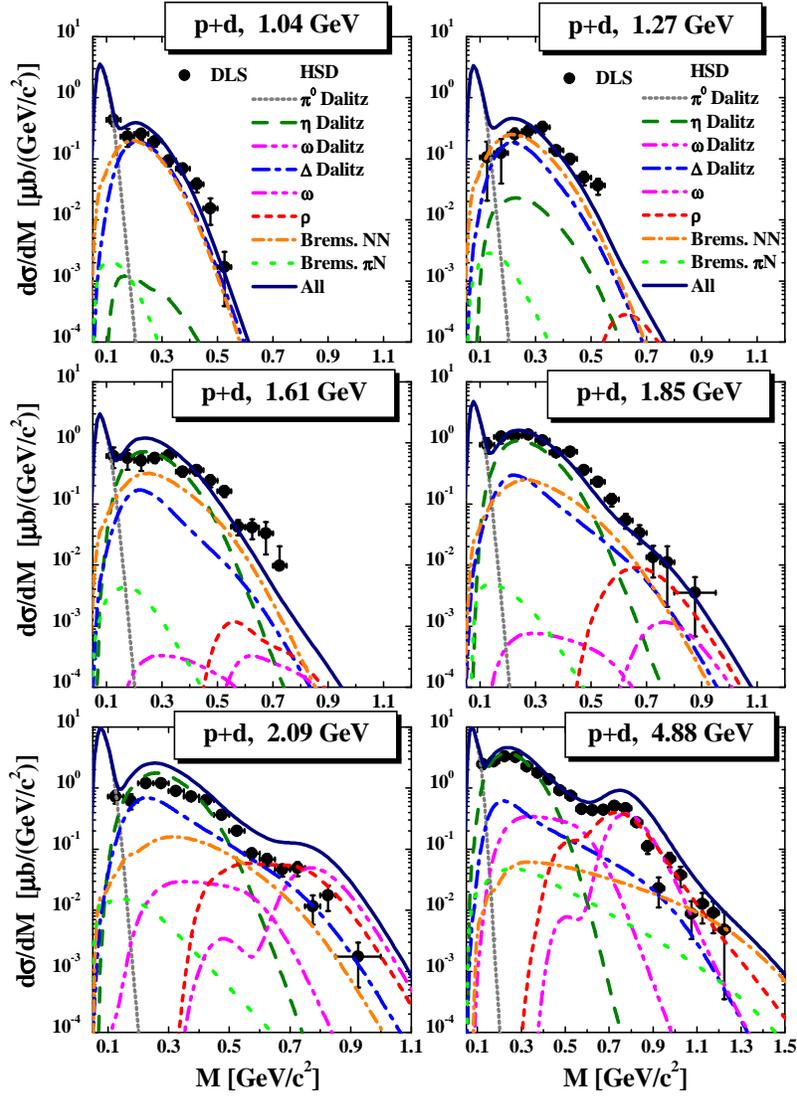,width=10.5cm}}
\caption{The differential cross section for $e^+e^-$ production in
$pd$ reactions  at bombarding energies of 1.04, 1.27, 1.61, 1.85,
2.09 and 4.88 GeV in comparison to the data from the BEVALAC
\cite{DLSpp} including the DLS acceptance filter and mass resolution
\cite{DLSpp}. The various contributions from the different channels
(in the HSD calculations) are indicated additionally (see color
coding in the legend). } \label{Fig10}
\end{figure}

The mass-differential dilepton cross section for $pd$ reactions is
compared in Fig. \ref{Fig10} to the BEVALAC data at the same
bombarding energies including the DLS acceptance filter and mass
resolution \cite{DLSpp}. In this case the relative contribution from
the various Dalitz decays and vector mesons decays stays about the
same as for $pp$ reactions except the contribution from
bremsstrahlung which at the lowest energies is of the same order
as the $\Delta$ Dalitz decay. Note that the $\rho$ and $\omega$
production is overestimated at 2.09 GeV as in the case of $pp$
reactions. Furthermore, coherent production of $\eta, \rho$ and
$\omega$ on the deuteron has been discarded in the HSD
calculations since the latter channels show up only very close to
the individual production thresholds and become negligible in $A+A$
reactions.

In view of the very limited DLS acceptance and mass resolution the
overall description of these 'elementary' reactions is acceptable
such that more complex reactions can be addressed.

\section{Dilepton production in nucleus-nucleus collisions}

A major aim of the HADES Collaboration was to clarify the
'DLS-puzzle', i.e. to verify/falsify the DLS data \cite{DLSnew}
from the experimental side. Accordingly, the same systems have
been reinvestigated at 1 A$\cdot$GeV with the HADES detector in
order to clarify the issue. The situation, however, is not as easy
since the DLS and HADES acceptances differ significantly
 \cite{DLSnew,HADES06}. However,
recently the HADES Collaboration has performed a direct comparison
with the DLS data for C+C at 1 A GeV \cite{DLSnew} by filtering the HADES data
with the DLS acceptance \cite{HADES07}. Both measurements
were found to agree very well \cite{HADES07}! Thus, the 'DLS
puzzle' has been solved from the experimental side; there is no
obvious contradiction between the DLS and HADES data! The major
goal for the transport models now is to verify the solution of the 'DLS-puzzle'
from the theoretical side. Note that transport calculations allow
for a direct comparison between the DLS and HADES measurements by
employing the different FILTER routines to the same set of calculated events.

\subsection{Comparison to the DLS data}

\begin{figure}[t]
\centerline{\psfig{figure=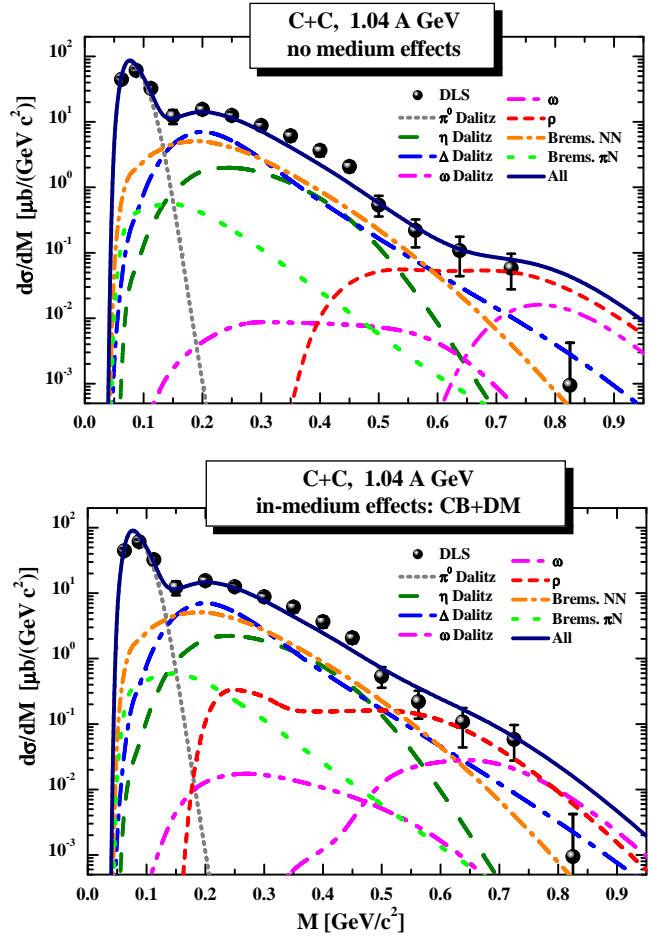,width=8.5cm}}
\caption{Results of the HSD transport calculation  for the mass
differential dilepton spectra in case of  $^{12}C+^{12}C$ at 1.04
A$\cdot$GeV in comparison to the DLS data \cite{DLSnew}. The
upper part shows the case of 'free' vector-meson spectral functions
while the lower part gives the result for the 'dropping mass +
collisional broadening' scenario. In both scenarios the DLS
acceptance filter and mass resolution have been incorporated. The
different color lines display individual channels in the transport
calculation (see legend). } \label{Fig11}
\end{figure}

We start with a reinvestigation of the DLS data employing the
novel cross sections for $\eta$ production, the modified $\Delta$
Dalitz decay contribution as well as the additional (and enhanced)
bremsstrahlung channels as described above. The results of our
transport calculation are displayed in Fig. \ref{Fig11} for
$^{12}C+^{12}C$ at 1.04 A$\cdot$GeV in case of 'free' vector-meson
spectral functions (upper part) and in case of the 'dropping mass +
collisional broadening' scenario (lower part) employing the DLS
acceptance filter and mass resolution.

In fact, the situation has substantially improved compared to the early
studies and the missing yield in the 'free' scenario is now reduced to
a factor of about 1.5 in the mass region from 0.25 to 0.5 GeV. This is
due to slightly higher contributions from $\Delta$ and $\eta$ Dalitz
decays and a significantly larger yield from bremsstrahlung channels.

To see the effects of these modifications quantitatively, we show
in Table 1 a comparison of our 'new' results with the 'old' ones
from Refs. \cite{BratRapp98,BratKo99}  -- indicated as 'new'/'old'
-- for $\eta$ and $\Delta$ Dalitz decays, $NN$ Bremsstrahlung
channels as well as for the sum of all contributions for C+C at 1
A GeV for invariant masses of 0.2, 0.35 and 0.5 GeV. Both results
correspond to the 'free' scenario.

\begin{table}[t]
\caption{The mass differential dilepton yield ${d\sigma\over dM}$
[$\mu$b/GeV] for $\eta$ and $\Delta$ Dalitz decays, $NN$
bremsstrahlung as well as the sum of all contributions for $C+C$
at 1 A GeV. The left numbers in each column stand for the present
results whereas the right numbers show the early results from
Refs. \protect\cite{BratRapp98,BratKo99} -- 'new'/'old'. The last
column presents the DLS experimental data \protect\cite{DLSnew}.
Both calculations refer to the 'free' scenario.}

\phantom{a}\vspace*{3mm}
\begin{center}
\begin{tabular}{ |c|c|c|c|c|c| }
\hline
M [GeV]   &    $\eta$      & $\Delta$     & Brems. NN     &       sum     &  exp. data   \\ \hline
0.2       & 1.75 / 1.3   & 7.05 / 5.5   & 5.0  / 1.88   & 14.2 / 4.7     &  $15.5 \pm     2.6$  \\
0.35      & 1.15 / 0.9   & 1.25 / 0.4   & 1.6  / 0.35   & 4.1     / 1.57  &  $6.1  \pm    1.1$  \\
0.5       & 0.12 / 0.11  & 0.16 / 0.03  & 0.22 / 0.04   & 0.54    / 0.28  &  $0.54 \pm    0.2$   \\
\hline
\end{tabular}
\end{center}
\phantom{a}\vspace*{3mm}
\end{table}

As seen from Fig. \ref{Fig11}, the bremsstrahlung yield now is similar
in shape and magnitude as the $\Delta$ Dalitz decay contribution and
enhanced by factor up to 5 due to the novel $NN$ bremsstrahlung cross
section from  \cite{Kaptari} and accounting for the contribution from
$pp$ bremsstrahlung additionally to $pn$.  The contribution of the
pion-nucleon bremsstrahlung is quite small in the C+C system due to the
limited energy available in meson-baryon collisions and the moderate
rescattering rate of pions.

As in our previous studies the $\eta$ Dalitz decay is subdominant.
We obtain only a slight enhancement (up to 30\%) of the $\eta$
yield at 1 A GeV compared to the early study \cite{BratRapp98}
since modifications of the $\eta$ production cross section in $NN$
reactions (cf. Section 3) are relevant at higher energies. The
enhancement of the high mass $\Delta$ Dalitz decay yield is
related mainly to the modifications in the LUND string model by
implementation of the mass dependent $\Delta$ width in the string
fragmentation (cf. Section 2.4) which enhances the population of
high mass $\Delta$ resonances. Additionally, we use now the
parametrization from Ernst et al. \cite{Ernst} for the $\Delta\to
N e^+e^-$ decay.

Medium modifications for the vector mesons turn out to yield an
enhancement in the region of 0.4 $< M <$ 0.5 GeV but are very moderate
due to the light system $C+C$. Though a slightly better
description of the DLS data is achieved it would be premature to
claim the presence of in-medium effects on the vector mesons from
these data.

\begin{figure}[t]
\centerline{\psfig{figure=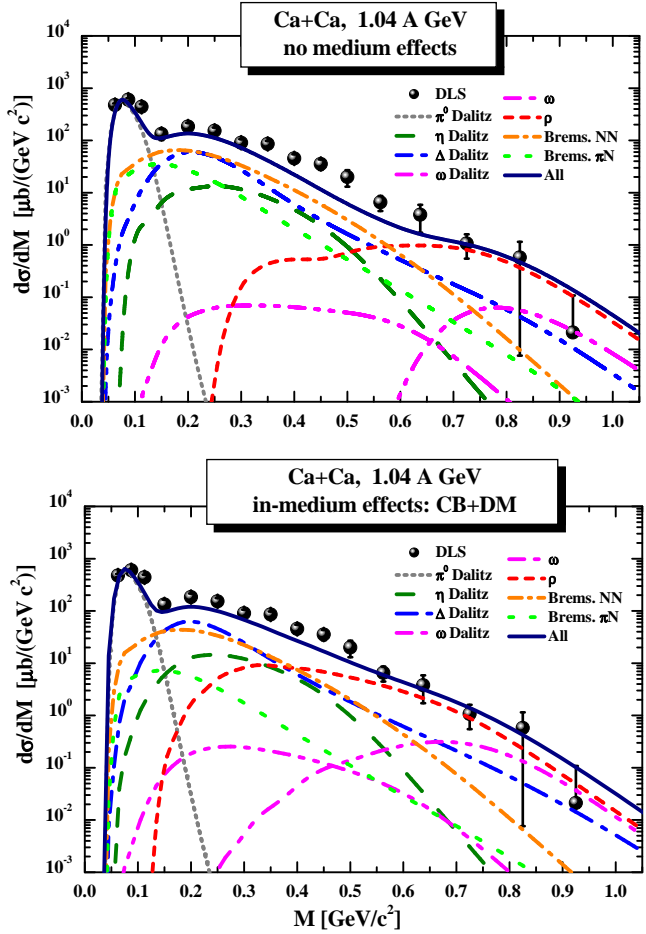,width=8.5cm}}
\caption{The mass differential dilepton spectra from HSD in case
of $^{40}Ca+^{40}Ca$ at 1.04 A$\cdot$GeV in comparison to the DLS
data \cite{DLSnew}. The upper part shows the case of 'free'
vector-meson spectral functions while the lower part gives the result
for the 'dropping mass + collisional broadening' scenario. In both
scenarios the DLS acceptance filter and mass resolution have been
incorporated. The different color lines display individual
channels in the transport calculation (see legend).}
\label{Fig12}
\end{figure}

Additional information is provided by the  $^{40}Ca+^{40}Ca$
system at 1.04 A$\cdot$GeV where the measured DLS spectra are
reproduced in a qualitatively and quantitatively similar manner by
the transport calculations (cf. Fig. \ref{Fig12}) as in case of
the lighter system. Again the combined bremsstrahlung channels
provide a contribution in the same order as the $\Delta$ Dalitz decay;
the $\eta$ Dalitz decay remains subdominant but the $\pi N$
bremsstrahlung increases compared to the $C+C$ system due to more
frequent pion rescattering on nucleons. In-medium effects for the vector
mesons can be indentified in the transport calculations but are
hard to see in the total dilepton spectra.

Is the 'DLS puzzle' solved? This question can only be answered in
a convincing manner by comparison with the recent HADES data.

\subsection{Comparison to HADES data}

\begin{figure}[!t]
\phantom{a}\vspace*{5mm}
\centerline{\psfig{figure=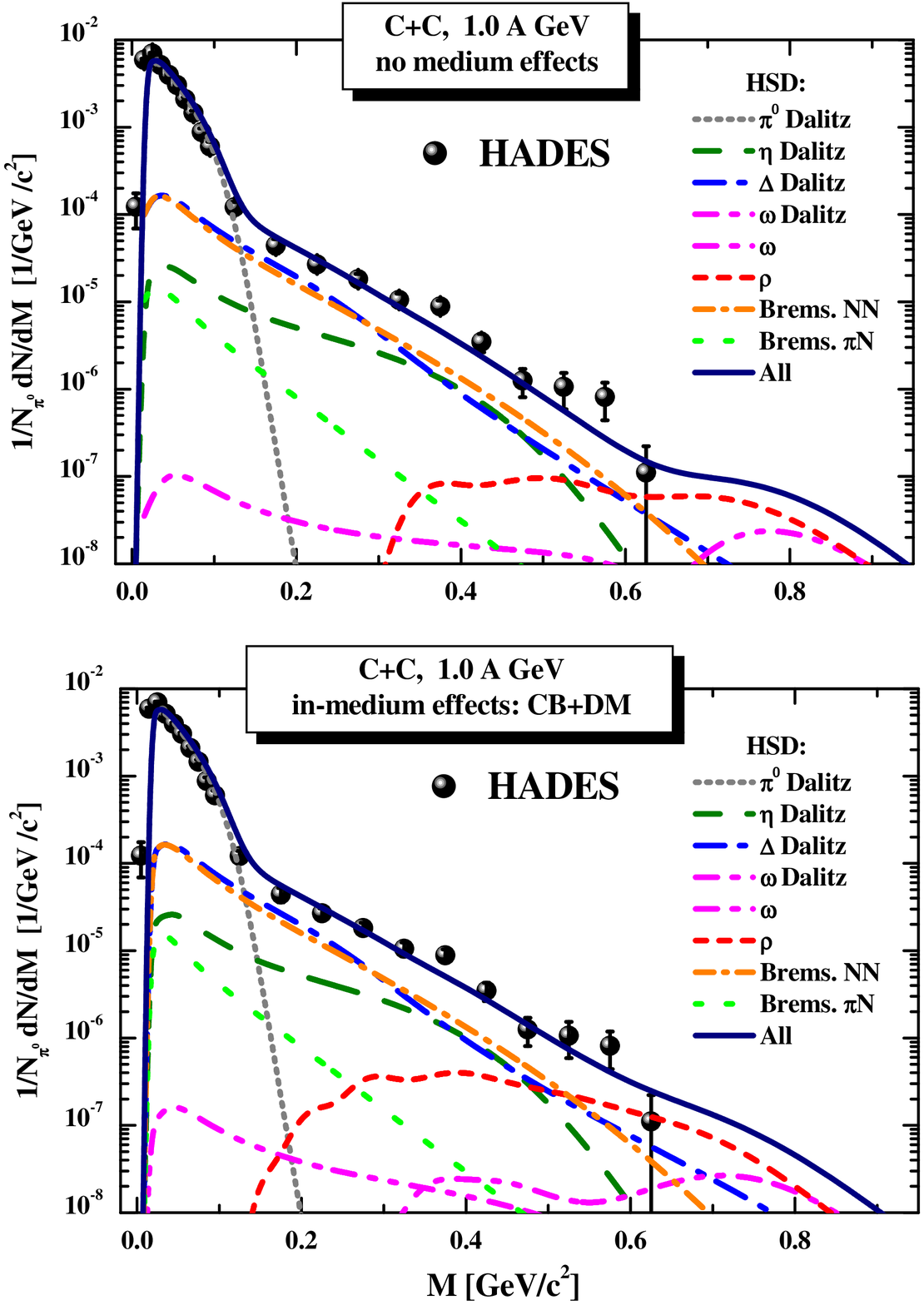,width=8.5cm}}
\caption{Results of the HSD transport calculation  for the mass
differential dilepton spectra - divided by the average number of
$\pi^0$'s - in case of $^{12}C+^{12}C$ at 1.0 A$\cdot$GeV in
comparison to the HADES data \cite{HADES07}. The upper part shows the case
of 'free' vector-meson spectral functions while the lower part gives
the result for the 'dropping mass + collisional broadening'
scenario. In both scenarios the HADES acceptance filter and mass
resolution have been incorporated. The different color lines
display individual channels in the transport calculation (see
legend). } \label{Fig13}
\end{figure}

The same multi-differential dilepton spectra - used for comparison
with the DLS data - are now filtered by the HADES acceptance
routines and smeared with the HADES mass resolution
\cite{HADES06}. A comparison of our calculations for the
$^{12}C+^{12}C$ system at 1.0 A$\cdot$GeV with the HADES data from
Ref. \cite{HADES07} is presented in Fig. \ref{Fig13} and
demonstrates that the agreement between data and transport
calculations is reasonable for the HADES data, too. In this case
the  differential dilepton spectrum is divided by the average
number of $\pi^0$'s which in experiment is determined by half of
the average number of ($\pi^+ +\pi^-$). The spectra look slightly
different in shape due to the much higher acceptance at low
invariant mass where the $\pi^0$ Dalitz decay practically exhausts
the dilepton spectrum. As in case of the DLS data the $\eta$
Dalitz decay turns out to be subdominant and the $\Delta$ Dalitz
decay to be of similar magnitude as the combined bremsstrahlung
contribution. Again the effect of in-medium vector-meson spectral
functions is hard to see in the total spectra.

\begin{figure}[t]
\phantom{a}\vspace*{5mm}
\centerline{\psfig{figure=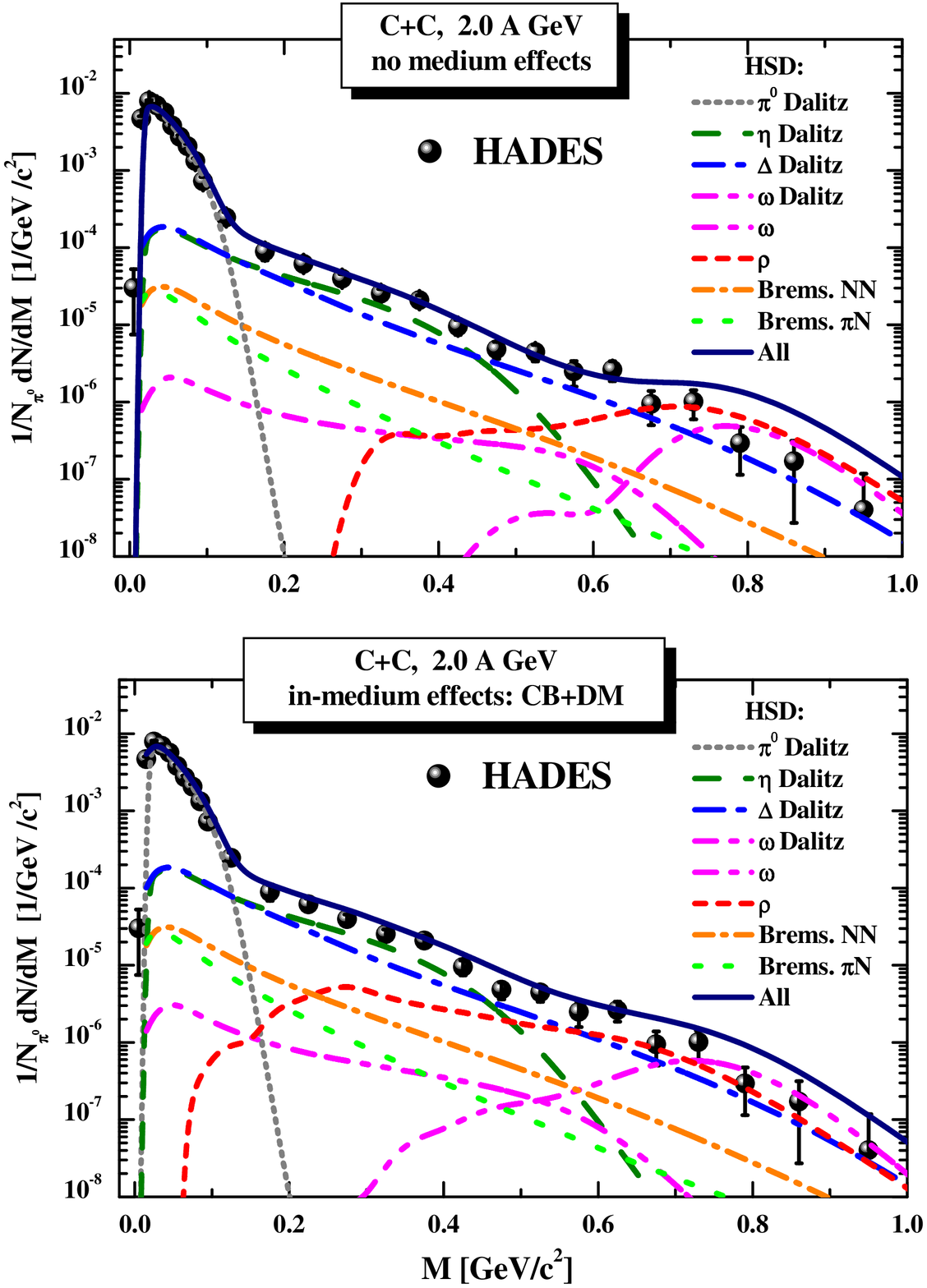,width=8.5cm}}
\caption{The mass differential dilepton spectra from HSD in case
of $^{12}C+^{12}C$ at 2 A$\cdot$GeV in comparison to the HADES
data \protect\cite{HADES06}. The upper part shows the case of 'free'
vector-meson spectral functions while the lower part gives the result
for the 'dropping mass + collisional broadening' scenario. In both
scenarios the HADES acceptance filter and mass resolution have
been incorporated. The different color lines display individual
channels in the transport calculation (see legend).} \label{Fig14}
\end{figure}

A comparison to the HADES mass-differential dilepton spectra for
$^{12}C+^{12}C$ at 2 A$\cdot$GeV \cite{HADES06} is presented in Fig.
\ref{Fig14} for the  'free' (upper part) and the in-medium scenario (lower part).
At the higher bombarding
energy the $\eta$ Dalitz decay provides the dominant contribution
in the mass region from 0.2 to 0.5 GeV followed by $\Delta$ Dalitz
decays and the combined bremsstrahlung channels. The mass region
around 0.75 GeV is overestimated in the 'free' scenario,
whereas including in-medium spectral functions for the vector mesons
the description of the data is improved (very similar to Ref.
\cite{wolfg2}) due a shifting of strength from the vector-meson
mass regime to lower invariant mass. However, the in-medium
effects for the light $C+C$ system are only very moderate.

\begin{figure}[h]
\phantom{a}\vspace*{5mm}
\centerline{\psfig{figure=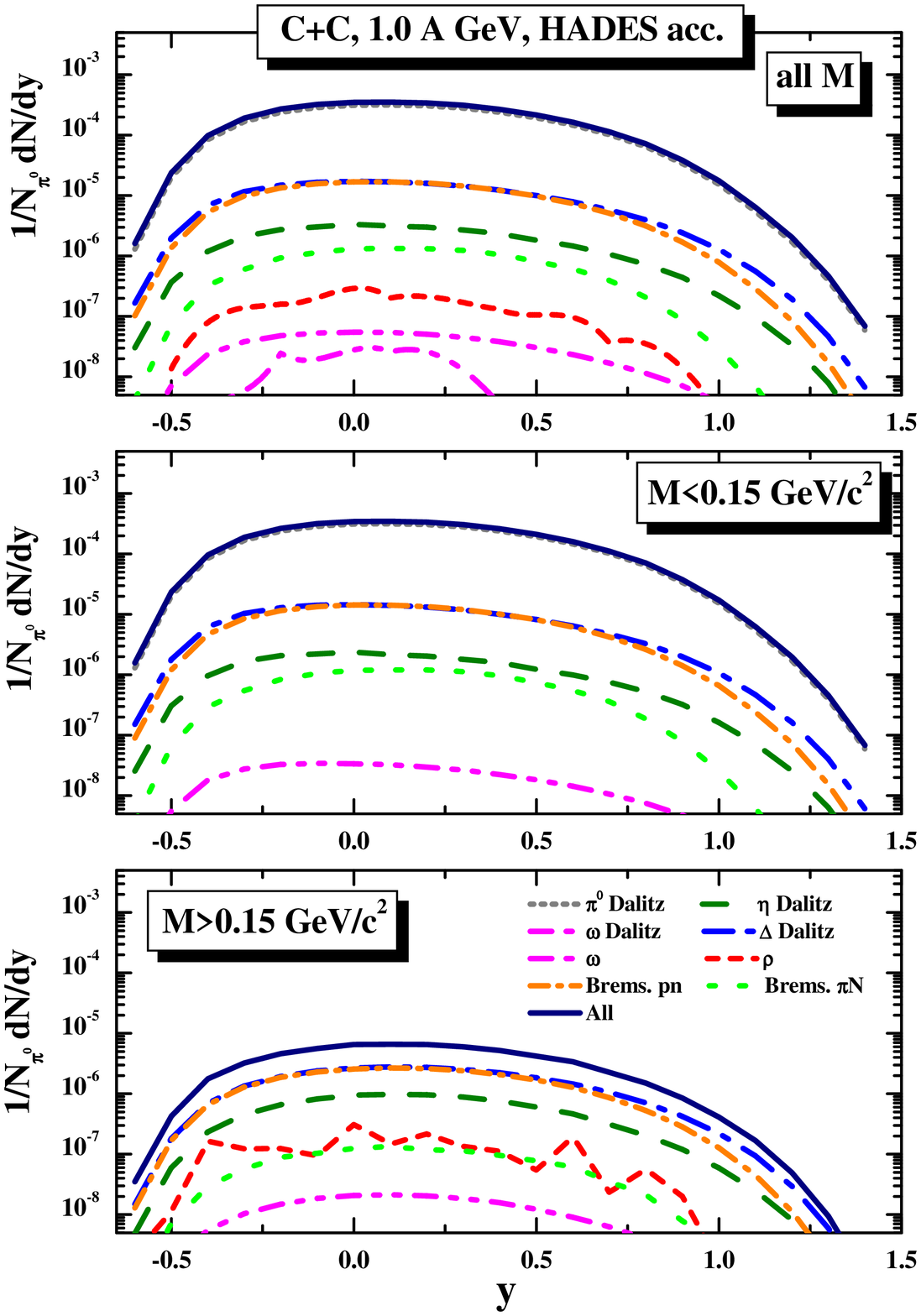,width=8.5cm}}
\caption{HSD predictions for the mass integrated rapidity
distributions for dileptons for $^{12}C+^{12}C$ at 1.0 A$\cdot$GeV
(top panel). A cut on low mass dileptons ($M <$ 0.15 GeV) is performed
in the middle part while the lower part shows the dilepton rapidity
distribution for a cut $M >$ 0.15 GeV. The individual colored lines
display the contributions from the various channels in the HSD calculations
(see color coding in the legend).} \label{Fig8}
\end{figure}

\begin{figure}[h]
\phantom{a}\vspace*{5mm}
\centerline{\psfig{figure=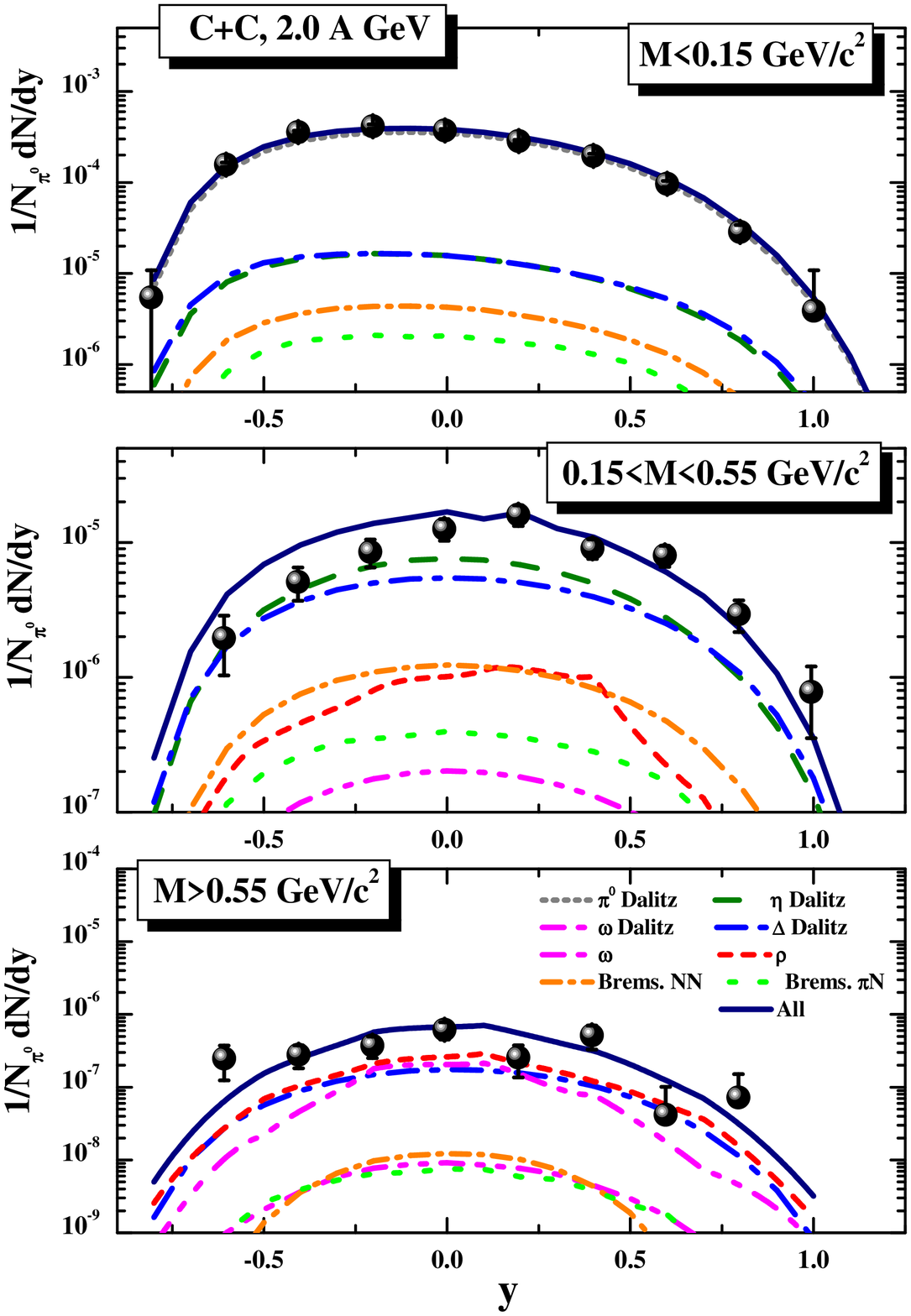,width=8.5cm}}
\caption{The mass integrated rapidity distributions for dileptons
from HSD for $^{12}C+^{12}C$ at 2.0 A$\cdot$GeV (top panel) in
comparison to the HADES data \protect\cite{HADES07pt}.
A cut on low mass dileptons ($M <$ 0.15 GeV) is performed in the upper
part, the middle panel corresponds to the cut
0.15 GeV $< M <$ 0.55 GeV while the lower part shows the
dilepton rapidity distribution for
$M >$ 0.55 GeV. The individual colored lines display the
contributions from the various channels in the HSD calculations (see
color coding in the legend).} \label{Fig8a}
\end{figure}

We now turn back to the system $^{12}C+^{12}C$ at 1.04 A$\cdot$GeV
and ask the question if the agreement in Fig. \ref{Fig13} might be
accidental? To address this issue we investigate the dilepton
rapidity spectra in the center-of-mass system for different cuts
on the invariant mass. The results of our
transport calculations are shown in Fig. \ref{Fig8} where the mass
integrated rapidity distributions are displayed on the upper
panel, a cut on low mass dileptons ($M <$ 0.15 GeV) is performed
in the middle part while the lower part shows the rapidity
distribution for a cut $M >$ 0.15 GeV. The individual colored
lines display the contributions from the various channels in the
HSD calculations. Since the experimental analysis for the dilepton
rapidity spectra (involving the same cuts) is still in progress we
still have to wait for the data to verify/falsify our predictions.

We recall that very similar investigations have been performed
for $^{12}C+^{12}C$ at 2.0 A$\cdot$GeV by the HADES Collaboration
\cite{HADES07pt}. Figure \ref{Fig8a} shows their mass integrated rapidity
distributions for dileptons from  $^{12}C+^{12}C$ at 2.0
A$\cdot$GeV (top panel) in comparison to the HSD calculations for
the in-medium scenario.  A cut on low mass dileptons ($M <$ 0.15
GeV) is performed in the upper part, the middle panel corresponds to the cut
0.15 GeV $< M <$ 0.55 GeV while the lower part shows
the dilepton rapidity distribution for  $M >$ 0.55 GeV. As
mentioned above, the $\eta$ Dalitz decay contribution becomes
dominant in the interval $0.15 < M< 0.55$ GeV at 2.0 A$\cdot$GeV,
whereas the contribution of bremsstrahlung decreases. In total we
find the HSD calculations to provide a good description of the
experimental rapidity distributions for all invariant mass ranges
considered.

\begin{figure}[h]
\phantom{a}\vspace*{5mm}
\centerline{\psfig{figure=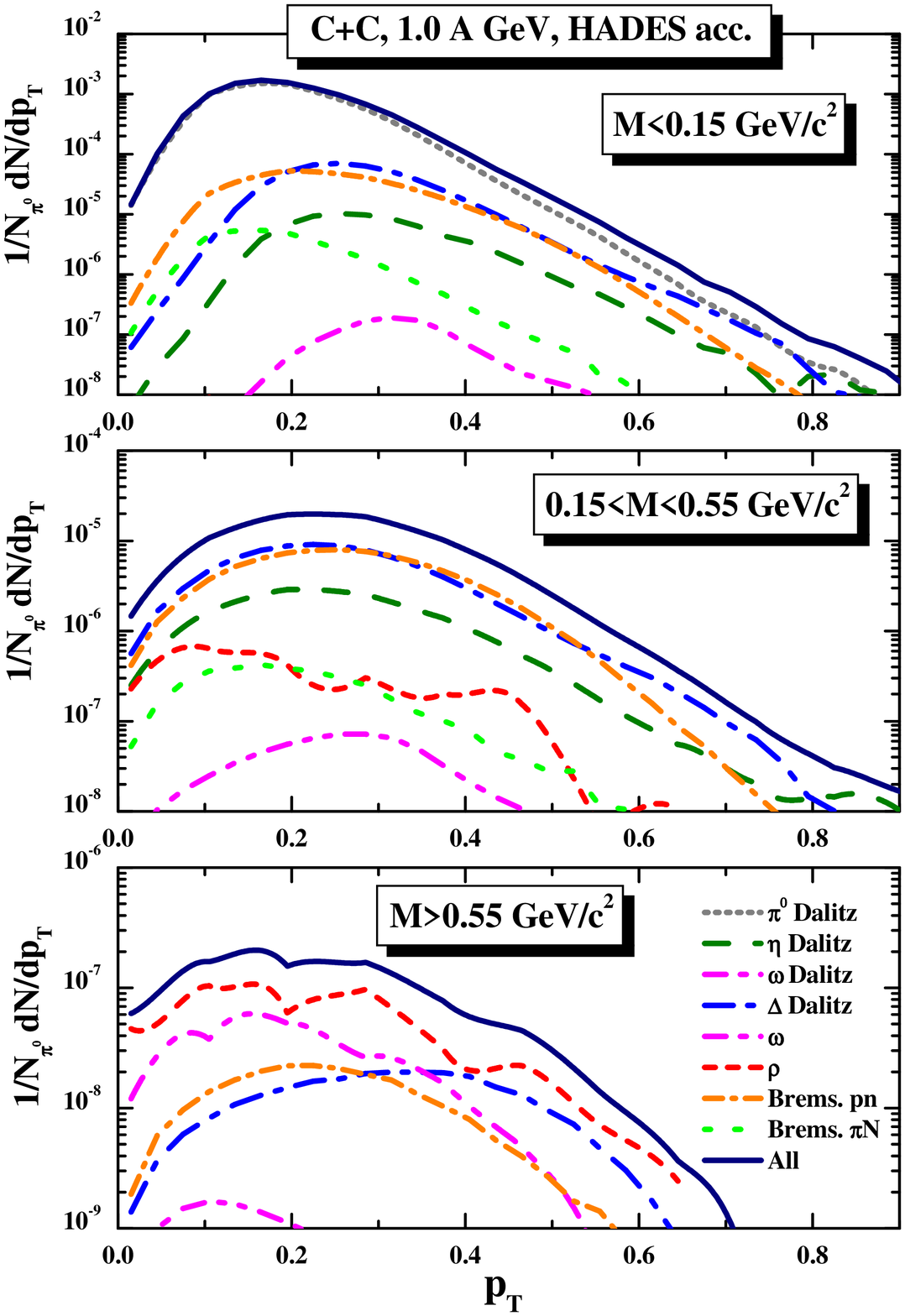,width=8.5cm}}
\caption{HSD predictions for the dilepton transverse momentum
spectra for $^{12}C+^{12}C$ at 1.0 A$\cdot$GeV.
The upper part corresponds to the mass range $M \leq$ 0.15 GeV,
the middle part to the interval  0.15 $\leq M \leq$ 0.55 GeV
and the lower part to $M \geq$ 0.55 GeV.
The individual colored lines display the
contributions from the various channels in the HSD calculations
(see color coding in the legend).} \label{Fig15}
\end{figure}

\begin{figure}[h]
\phantom{a}\vspace*{5mm}
\centerline{\psfig{figure=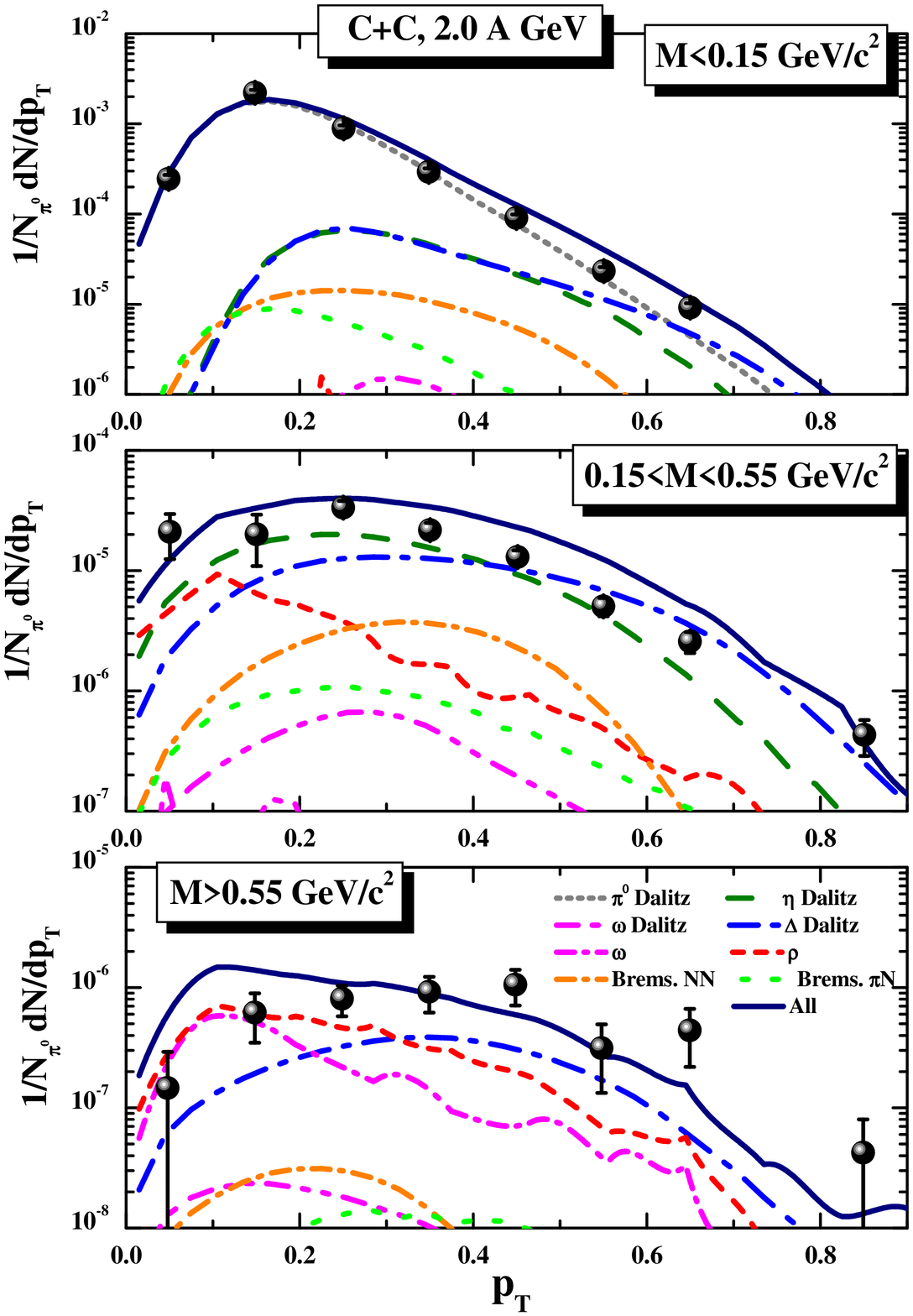,width=8.5cm}}
\caption{The dilepton transverse momentum spectra from HSD
for $^{12}C+^{12}C$ at 2.0 A$\cdot$GeV in
comparison to the data from the HADES Collaboration \protect\cite{HADES07pt}.
The upper part corresponds to the mass range $M \leq$ 0.15 GeV,
the middle part to the mass range 0.15 GeV $\leq M \leq$ 0.55 GeV
and the lower part  to $M \geq$ 0.55 GeV.
The individual colored lines display the
contributions from the various channels in the HSD calculations
(see line coding in the legend).} \label{Fig16}
\end{figure}

Apart from dilepton rapidity distributions (for different mass bins)
also transverse momentum spectra of $e^+e^-$ pairs provide additional
information.  For comparison with future HADES data we show the results
of our HSD calculations in Fig. \ref{Fig15} for the dilepton transverse
momentum spectra in the mass range $M <$ 0.15 GeV (upper part), 0.15
$\leq M \leq$ 0.55 GeV (middle part) and $M >$ 0.55 GeV (lower part)
for $^{12}C+^{12}C$ at 1.0 A$\cdot$GeV. For the lowest invariant mass
bin the transverse momentum spectrum is entirely dominated by the
$\pi^0$ Dalitz decays, however, followed by the bremsstrahlung
contribution at low $p_T$. In the interesting mass range 0.15 $< M < $
0.55 GeV the bremsstrahlung and $\Delta$ Dalitz contributions dominate
which might provide a handle to establish the strength of the
bremsstrahlung channels independently from the mass differential
spectra.

Whereas the data analysis for the dilepton $p_T$ spectra for
$^{12}C+^{12}C$ at 1.04 A$\cdot$GeV is still in progress we may gain
related information in comparison to HADES data at 2.0 A$\cdot$GeV.
Accordingly, the dilepton transverse momentum spectra from the
in-medium HSD calculations for $^{12}C+^{12}C$ at 2.0 A$\cdot$GeV are
shown in Fig. \ref{Fig16} in comparison to the data from the HADES
Collaboration \protect\cite{HADES07pt}. The upper part corresponds to
the mass range $M \leq$ 0.15 GeV/c$^2$, the middle part to 0.15 $\leq M
\leq$ 0.55 GeV/c$^2$ and the lower part to $M \geq$ 0.55 GeV/c$^2$.
Indeed, we find again a quite reasonable agreement with the HADES data
on the transverse momentum distributions for practically all mass
intervals. A slight overestimation of the experimental data at lower
$p_T$ is seen for the mass interval $M \geq$ 0.55 GeV/c$^2$ (lower part
of Fig. \ref{Fig16}). This is due to the fact that the 'dropping mass +
collisional broadening' scenario for the vector mesons leads to a shift
of the $p_T$ spectra to the low $p_T$ region. This effect is stronger
for  heavy systems with larger baryon densities and might be used to
distinguish the different in-medium scenarios from the experimental
side.

\subsection{Predictions at SIS energies}

The DLS data on $e^+e^-$ production from elementary reactions like
$pp$ and $pd$ collisions have been taken in a rather small range
of phase space with limited statistics. This is sufficient for
'pilot experiments' but decisive measurements - allowing for a
distinction between different model scenarios - are still missing.
To this end the HADES Collaboration is in the process of
performing experiments with $pp$ and/or $pd$ reactions at GSI with
improved statistics, mass resolution and phase-space coverage.
These new experiments will also shed further light on the size of
the bremsstrahlung channels in competition with hadron Dalitz
decays. In order to prove/disprove the interaction scenarios
employed here in HSD we present predictions for the future
measurements within the presently known HADES detector
acceptance\footnote{Any change in experimental cuts or in the data
analysis in future will easily be taken into account since the HSD
'raw data' will be made available to the HADES Collaboration.}.

\begin{figure}[t]
\phantom{a}\vspace*{5mm}
\psfig{figure=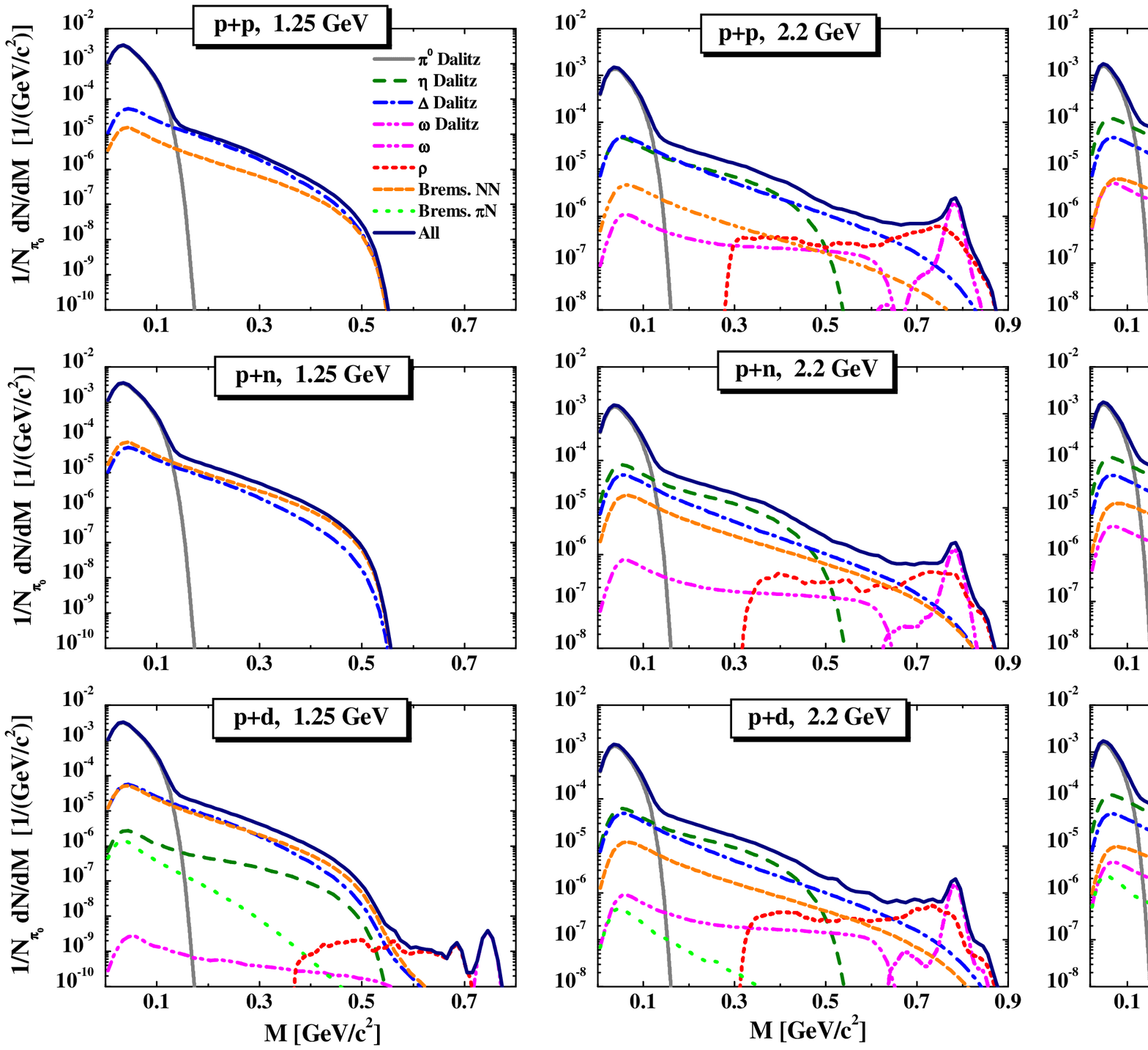,width=11cm}
\caption{The differential cross section for $e^+e^-$ production in
$pp$ (upper part), $pn$ (middle part) and $pd$ (lower part)
reactions  at bombarding energies of 1.25 GeV (l.h.s.), 2.2 GeV
(middle) and 3.5 GeV (r.h.s.) including the HADES acceptance
filter and  mass resolution (see legend for the different color
coding of the individual channels). } \label{Fig17}
\end{figure}

We start with elementary reaction channels: Figure \ref{Fig17} shows
the differential cross section for $e^+e^-$ production in $pp$ (upper
part), $pn$ (middle part) and $pd$ (lower part) reactions at bombarding
energies of 1.25 GeV (l.h.s.), 2.2 GeV (middle) and 3.5 GeV (r.h.s.)
including the actual HADES acceptance filter and mass resolution (that
differs in the magnetic settings for the different systems and bombarding energy).
We note that the dilepton spectra are normalized to the number
of $\pi^0$ taken from the HSD calculations and not to $(\pi^+ +\pi^-)/2$
as for the heavy-ion reactions. This is important to point out due
to the strong isospin asymmetry in $pp, pn$ and $pd$ reactions.

It is seen from Fig. \ref{Fig17} that for low
invariant masses ($< 0.15$ GeV) the $\pi^0$ Dalitz decays
dominates which can be controlled additionally by direct pion
measurements. At 1.25 GeV the $\Delta$ Dalitz decay (dash-dotted
blue line) is by far the dominant channel in $pp$ collisions
whereas the $NN$ bremsstrahlung channels (dash-dotted orange line)
overtake for $pn$ collisions at this energy. In $pd$ reactions the
$\Delta$ Dalitz decay and the bremsstrahlung channels contribute
with roughly the same magnitude in the entire invariant mass
range. The $\eta$ Dalitz decay slightly contributes in the $pd$
case due to the internal Fermi motion of the nucleons in the
deuteron. Discarding the low mass $\pi^0$ decays we find a
dominance of $\eta$ and $\Delta$ Dalitz decays at 2.2 and 3.5 GeV for
$pp$, $pn$ and $pd$ collisions up to invariant masses of 0.6 GeV.
Bremsstrahlung channels here are lower by more than an order of
magnitude and are hardly to access at this energy. On the other
hand the $\pi^0$, $\eta$ and $\Delta$ Dalitz decays may be
controlled by direct hadronic measurements.
In the vector-meson mass range ($\sim$ 0.8 GeV) the
dilepton yield is clearly dominated by the direct $\rho$ and
$\omega$ decays; a separation of these channels is possible by
measuring additionaly the $\pi^0 \gamma$ decay channel of the
$\omega$ meson as in Ref. \cite{tapselsa}. Note that the
contribution from pion rescattering on the spectator nucleon in
case of $pd$ reactions (dotted light green lines) is down by
roughly two orders of magnitude and will be hard to identify
experimentally.

\begin{figure}[t]
\phantom{a}\vspace*{5mm}
\centerline{\psfig{figure=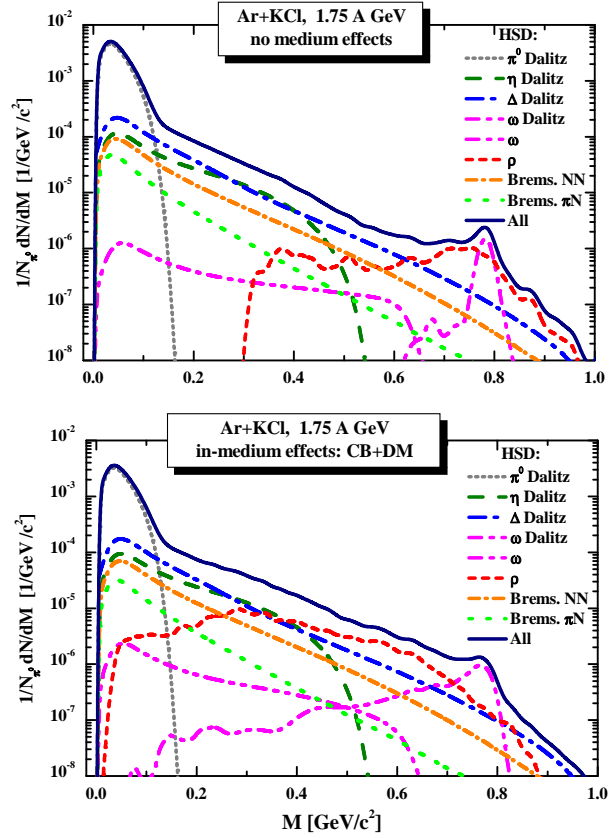,width=8.5cm}}
\caption{The HSD predictions  for the mass differential dilepton
spectra - divided by the average number of $\pi^0$'s taken
as $(\pi^+ +\pi^-)/2$ - in case of $Ar+KCl$ at 1.75 A$\cdot$GeV. The
upper part shows the case of 'free' vector-meson spectral functions
while the lower part gives the result for the 'dropping mass +
collisional broadening' scenario. In both scenarios the actual HADES
acceptance filter and mass resolution have been incorporated (see
legend for the different color coding of the individual channels). }
\label{Fig18}
\end{figure}

We continue with nucleus-nucleus collisions. Since the
modifications of the $\rho$-meson spectral function only leave
moderate traces in the differential dilepton spectra for
$^{12}C+^{12}C$ (cf. Section 5.2) we provide predictions for
$Ar + KCl$ at 1.75 A$\cdot$GeV, a system which is currently analyzed by
the HADES Collaboration and might allow for a closer view on the
$\rho$-meson properties in a denser medium. The HSD predictions
for the mass differential dilepton spectra - divided by the
average number of $(\pi^+ +\pi^-)/2$ - in this case  are presented in Fig.
\ref{Fig18}. The upper part shows the case of 'free' vector-meson
spectral functions while the lower part gives the result for the
'dropping mass + collisional broadening' scenario. In both
scenarios the actual HADES acceptance filter and mass resolution
have been incorporated. Since the standard Dalitz decays are
roughly the same in both scenarios let us concentrate on the
direct vector meson decay channels, i.e. the dashed red lines for
the $\rho$ meson and the double-dot - dashed purple lines  for the
$\omega$ meson. The effect of the finite mass resolution can
clearly be seen in case of the $\omega$ meson for the 'free'
scenario. The peak appears at the pole mass and is dominantly
smeared by the experimental mass resolution (except from the
moderate vacuum decay width).

In the 'dropping mass + collisional broadening' scenario the shift
of the $\omega$ strength to lower invariant masses is clearly seen
in the calculations but will be very hard to access in experiment.
In this respect it might be more favorable to look for the $\pi^0
\gamma$ decay  in comparison to the $e^+e^-$ decay in order to
observe experimentally in-medium effects for the $\omega$ meson.
In case of the $\rho$ meson the effects of a dropping mass +
collisional broadening become clearly visible for invariant masses
above 0.3 GeV resulting in a significant enhancement (by more than
a factor of two) in the mass differential dilepton spectra.
Actually this in-medium enhancement is slightly lower than
predicted in the (more simplified) on-shell quasi-particle
approach in the nineties \cite{CBRW97,CBRep98,BratRapp98,BratKo99}
but should be measured experimentally in
case of high statistics.

\section{Summary}

In this study a detailed analysis of dilepton production in $C+C$
and $Ca+Ca$ at 1-2 A$\cdot$GeV as well as in elementary $pp$ and
$pd$ reactions at 1-5 GeV has been presented within the
microscopic HSD transport approach that incorporates a full
off-shell propagation of the vector mesons in line with Refs.
\cite{Cass_off1,Cass_off2}. A comparison of dilepton spectra from
'elementary' $pp$ and $pd$ reactions with the DLS data in the
energy range from 1 to 4.9 GeV specifies the isospin dependence of
the various channels and shows the adequacy/deficiency of the
parametrization or model assumptions in the HSD approach.

Our present HSD model includes the following novel developments
with respect to the previous studies a decade ago
\cite{CBRep98,CBRW97,BratRapp98,BratKo99}:
\begin{itemize}
\item off-shell dynamics for vector mesons and explicit (nonperturbative)
propagation of vector mesons also at SIS energies
\item improved cross sections for $\eta$ production in $pp$ and $pn$
reactions
\item extension of the LUND string model for the production of particles
according to  'in-medium' spectral functions
\item (enhanced) bremsstrahlung contributions according to
Ref. \cite{Kaptari} which proofs
vital for a solution of the DLS puzzle.
\end{itemize}

Furthermore, we have analyzed the $\pi^0$ and $\eta$ production in
$C+C$ collisions at SIS energies.  The HSD results for $\pi^0$ and
$\eta$ productions have been found to be in a good agreement with
the data from the TAPS Collaboration from 0.8 to 2 A$\cdot$ GeV.
These studies pin down the dominant contributions from $\pi^0$ and
$\eta$ Dalitz decays in the low mass dilepton spectra and also to
a large extent the contribution from $\Delta$ Dalitz decays which
are closely linked to the pion multiplicity.

On the vector meson side different scenarios of in-medium
modifications of $\rho$'s and $\omega$'s, such as collisional
broadening and additionally dropping vector meson masses (cf. Fig.
\ref{Fig0}), have been investigated and the possibilities for an
experimental observation of in-medium effects in $AA$ reactions
has been discussed. We find that within the extended off-shell
transport approach we achieve a simultaneous description of the
differential dilepton spectra for $^{12}C+^{12}C$ at 1.04 A$\cdot$
GeV from the DLS Collaboration as well as the HADES Collaboration
which points towards a consistency of the independent measurements
within error bars. In general the experimental spectra are better
described by the in-medium scenarios for the vector mesons,
however, the modifications of the spectra relative to the 'free'
scenario are very moderate for the light $C+C$ systems.

In order to shed some light on perspectives for heavier systems we
have provided detailed predictions for $Ar + KCl$ at 1.75
A$\cdot$GeV as well as  elementary $pp$ and $pd$ reactions at 1.25
GeV, 2.2 GeV and 3.5 GeV. The latter systems are addressed also by the
HADES Collaboration and in future will shed light on the
contribution of bremsstrahlung channels in elastic and inelastic
channels at about 1 A$\cdot$GeV and
the in-medium modifications of the $\rho$-meson at 1.75
A$\cdot$GeV in the heavier nucleus-nucleus system. We stress that
new calculations of
dilepton bremsstrahlung - based on a consistent full $G-$ (or
$T-$) matrix approach - are urgently needed from the theoretical side
in order to clarify the presently open situation.

\section*{Acknowledgement}
The authors acknowledge inspiring discussions with T. Galatyuk,
R. Holtzmann, V. Metag, Yv. Pachmayer, P. Salabura, J. Stroth
and M. Sudol.


\end{document}